%% file: main.tex
\documentclass[journal]{IEEEtran}
\usepackage[colorlinks, citecolor=blue]{hyperref}

\usepackage{cite}
\usepackage{xcolor}
\usepackage{array}
\usepackage{url,hyperref,booktabs}
\usepackage{pifont}
\usepackage{makecell}
\usepackage{arydshln}
\usepackage{tablefootnote}

\ifCLASSINFOpdf
  \usepackage[pdftex]{graphicx}
  \DeclareGraphicsExtensions{.pdf,.jpeg,.png}
  \usepackage{epstopdf}
  \usepackage{subcaption}
\else
  \usepackage[dvips]{graphicx}
\fi
\usepackage{amsmath}
\usepackage{amssymb}
\usepackage{multirow}
\usepackage[switch]{lineno}
\usepackage{array}

\usepackage{stfloats}
\usepackage[symbol]{footmisc}

\makeatletter
\def\endthebibliography{%
  \def\@noitemerr{\@latex@warning{Empty `thebibliography' environment}}%
  \endlist
}
\makeatother

\begin{document}
%
\title{Exploring Length Generalization For Transformer-based Speech Enhancement}
%
%
%

\author{Qiquan~Zhang,~\IEEEmembership{Member,~IEEE,}
        Hongxu Zhu,~\IEEEmembership{Member,~IEEE,}
        Xinyuan Qian,~\IEEEmembership{Member,~IEEE,}
        Eliathamby Ambikairajah,~\IEEEmembership{Life Senior Member,~IEEE,} 
        and Haizhou Li,~\IEEEmembership{Fellow,~IEEE}
\thanks{Manuscript received date; revised date. 
This work is supported in part by Shenzhen Science and Technology Program (Shenzhen Key Laboratory, Grant No. ZDSYS20230626091302006), in part by Shenzhen Science and Technology Research Fund (Fundamental Research Key Project, Grant No. JCYJ20220818103001002), in part by Program for Guangdong Introducing Innovative and Enterpreneurial Teams, Grant No. 2023ZT10X044, and in part by Australian Research Council (ARC) Discovery Grant DP210101228. \textit{(Corresponding author: Hongxu Zhu)}}
\thanks{Qiquan Zhang and Eliathamby Ambikairaiah are with the School of Electrical Engineering and Telecommunications, The University of New South Wales, Sydney, NSW 2052, Australia (e-mail: {zhang.qiquan}@outlook.com; {e.ambikairajah}@unsw.edu.au).}
\thanks{Hongxu Zhu is with the Department of AI System, Fano, Hong Kong, HKG, China (e-mail: {hongxuzhu2-c}@my.cityu.edu.hk).}
\thanks{Xinyuan Qian is with the School of Computer and Communication Engineering, University of Science and Technology Beijing, 100083, China (e-mail: {qianxy}@ustb.edu.cn)}
\thanks{Haizhou Li is with the School of Data Science, The Chinese University of Hong Kong, Shenzhen, Guangdong 518172, China (e-mail: {haizhouli}@cuhk.edu.cn).
}

}

%
%

\markboth{Journal of \LaTeX\ Class Files,~Vol.~14, No.~8, August~2019}%
{Shell \MakeLowercase{\textit{et al.}}: Bare Demo of IEEEtran.cls for IEEE Journals}
%



\maketitle


\begin{abstract} 

Transformer network architecture has proven effective in speech enhancement. However, as its core module, self-attention suffers from quadratic complexity, making it infeasible for training on long speech utterances. In practical scenarios, speech enhancement models are often required to perform on noisy speech at run-time that is substantially longer than the training utterances. It remains a challenge how a Transformer-based speech enhancement model can generalize to long speech utterances. In this paper, extensive empirical studies are conducted to explore the model's length generalization ability. In particular, we conduct speech enhancement experiments on four training objectives and evaluate with five metrics. Our studies establish that positional encoding is an effective instrument to dampen the effect of utterance length on speech enhancement. We first explore several existing positional encoding methods, and the results show that relative positional encoding methods exhibit a better length generalization property than absolute positional encoding methods. Additionally, we also explore a simpler and more effective positional encoding scheme, i.e. LearnLin, that uses only one trainable parameter for each attention head to scale the real relative position between time frames, which learns the different preferences on short- or long-term dependencies of these heads. The results demonstrate that our proposal exhibits excellent length generalization ability with comparable or superior performance than other state-of-the-art positional encoding strategies.

\end{abstract}


%
\IEEEpeerreviewmaketitle

\section{Introduction}\label{sec:1}
%
%
%
%



\IEEEPARstart{I}{n} \textcolor{black}{our daily living environments, speech signals are often corrupted by background noise. This can adversely affect the performance of speech applications, such as assistive listening devices, mobile speech communication, speech coding, and automatic speech recognition (ASR). One popular way to mitigate this issue is to employ a speech enhancement front-end to improve speech intelligibility and quality. Monaural speech enhancement, which recovers the clean speech from a noisy recording, is one of the examples.}


\begin{figure}[!ht]
\centering
\centerline{\includegraphics[width=0.81\columnwidth]{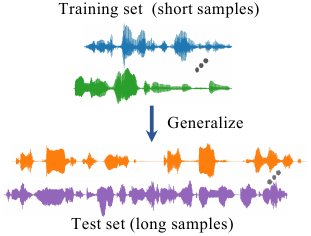}}
\caption{\textcolor{black}{Length generalization in speech enhancement: The ability to learn from short speech samples (training set) to generalize to longer speech samples (test set).} 
}
\label{generalize}
\vspace{-0.5em}
\end{figure}

\textcolor{black}{Monaural speech enhancement has been an active research direction in speech signal processing for decades, leading to the development of a variety of methods. Conventional speech enhancement techniques \cite{loizou} mainly involve spectral subtraction \cite{boll1979}, statistical model-based algorithms \cite{mmse,mmse2017,zhang2019}, and Wiener filtering \cite{wiener1996}. However, due to certain restrictions and simplifications about speech and noise characteristics, these algorithms often fail to eliminate fast-varying noise signals \cite{loizou}. In recent decades, facilitated by large-scale training data and powerful computing resources, deep learning technology has promoted enormous progress in speech enhancement \cite{overview2018,over2022}, demonstrating substantial performance improvements over conventional techniques.}

\textcolor{black}{Transformer network architectures have been widely used in speech and language processing. 
They have shown state-of-the-art performance in many speech processing tasks, such as speech enhancement \cite{cleanunet,mhanet,restcnsa,tgsa,zhao2020,ripple} and speech recognition \cite{conformer-asr}. The multi-head self-attention (MHSA) module is a core component in Transformers and explicitly learns the relevance of each time frame in a sequence with respect to all other time frames, generating context-informed feature representations. This enables Transformers to capture long-range temporal dependencies more efficiently.}

The generalization capability to unseen conditions during training is crucial for supervised learning tasks. In speech enhancement, generalization to unseen noise, speakers, and signal-to-noise ratios (SNRs) have been extensively studied. Besides the three aspects, the speech enhancement model is expected to generalize to perform on speech utterances significantly longer than those (usually with a fixed maximum length) seen during training. Self-attention suffers from the computation cost quadratic with respect to the sequence length, which makes it infeasible to use sufficiently long speech utterances for training Transformer. A natural question arises: ``Can the Transformer-based speech enhancement model effectively learn from shorter speech samples and generalize to longer ones?'', which is recently defined as the problem of length generalization \cite{anil2022exploring}. Aside from mitigating the computation cost issue, length generalization is an intriguing theoretical problem. 
In this paper, we extend our previous work~\cite{zhang2024exploration} by addressing issues how Transformer-based speech enhancement models handle long noisy speech samples. In Fig.~\ref{generalize}, we illustrate the concept of length generalization in speech enhancement.

In the vanilla Transformer formula~\cite{attention2017}, a sinusoidal position embedding is used to allow the model to handle sequences of variable length between training and testing. However, we observe that such position embedding is not effective. Studies have shown that relative position encoding (RPE) schemes exhibit a better generalization property, e.g., T5-Bias~\cite{T5}, TISA~\cite{tisa}, and KERPLE~\cite{kerple}. We are inspired by RPE to enable Transformer-based speech enhancement for length generalization at run-time.

\textcolor{black}{In this paper, we explore seven existing positional encoding methods in length generalization.} 
These methods are initially designed for natural language processing. Note that speech spectrogram has a more rigid temporal order than text tokens in natural language. We argue that the complex formulations for positional encoding in T5-Bias, KERPLE, and TISA may not be effective for speech tasks. Hence, we attempt to explore a more efficient RPE scheme with a simpler design, i.e., LearnLin, which encodes position information by exploiting only one learnable parameter to explicitly scale the relative distance between different time frames. A different parameter is learned for each attention head, which adjusts the preferences of different heads on short- or long-range temporal dependencies. It biases the raw attention scores with summation operation before Softmax normalization. \textcolor{black}{In summary, this paper extends the previous work~\cite{zhang2024exploration} with the following contributions:}

\begin{itemize}
    \item \textcolor{black}{We explore a simpler and more efficient positional encoding scheme (LearnLin) for Transformer speech enhancement, which employs only one trainable parameter in each head to scale the relative distance to learn the preferences on long- or short-term dependencies. Experiments show that LearnLin is superior or on par with other prior models in terms of length generalization.}
    
    \item \textcolor{black}{We conduct a more comprehensive and systematic exploration of seven positional encoding approaches on four training targets (including both mapping- and masking-based methods) to dampen the effect of utterance length, across both causal and non-causal model configurations. 
    }
    \item \textcolor{black}{We further evaluate the performance under chunk-based processing, both without overlap and with a 50\% overlap. The results highlight the effectiveness and importance of performing length generalization.}
\end{itemize}

\textcolor{black}{The remainder of this paper is structured as follows. In Section~\ref{sec:2}, we describe related works, including speech enhancement and length generalization. Section~\ref{sec:3} elaborates on the positional encoding methods. Section~\ref{sec:4} formulates the research problem. \textcolor{black}{In Section~\ref{sec:5}, we describe speech enhancement with position-aware Transformer Section~\ref{sec:6} describes the experimental setup. The experimental results are presented in Section~\ref{sec:7}. Finally, Section~\ref{sec:8} concludes this study.}}


\section{Related Work} \label{sec:2}
\textcolor{black}{Neural speech enhancement schemes can be grouped into waveform-domain methods and time-frequency (T-F) domain methods. The waveform-domain methods recover the target speech in an end-to-end manner by optimizing a deep neural network (DNN) model to directly predict the raw waveform of the clean speech from the noisy one~\cite{SEGAN,2017raw, kolbaek2020loss,2018raw,demcus,cleanunet}. Many other speech enhancement techniques are carried out in a time-frequency (T-F)  domain~\cite{overview2018,over2022}, such as spectral mapping and T-F masking methods. Spectral mapping methods optimize the DNN model to map the T-F representation (or spectral feature) of the noisy speech to the clean T-F representation. The most commonly used spectral features for spectral mapping include log-power spectrum (LPS)~\cite{yongxu2015}, magnitude spectrum (MS)~\cite{crn2018,GRN,restcnsa,sa-tcn}, and complex spectrum (CS)~\cite{2018complex,tan2019learning}. In T-F masking methods, the DNN model is trained to predict a T-F mask, which is then applied to the noisy T-F representation to obtain the clean one. Furthering the idea of auditory T-F masking, the ideal binary mask (IBM) is proposed as the training objective to accomplish speech enhancement, where the speech- and noise-dominated T-F units are labeled as one and zero, respectively~\cite{wang2013}. Later, different types of T-F masks are developed along this line of research, where the most popular ones include the spectral magnitude mask (SMM)~\cite{targets}, ideal ratio mask (IRM) \cite{targets}, complex ideal ratio mask (cIRM)~\cite{cirm}, and phase-sensitive mask (PSM)~\cite{psm}.}

\textcolor{black}{
Initial speech enhancement models typically employ a feed-forward DNN (FNN) as the backbone network, which can only leverage limited temporal correlations provided by a sliding window of consecutive time frames~\cite{yongxu2015,targets,cirm}. Later, with the capability to effectively exploit the long-range temporal dependencies present in speech signals, the long short-term memory (LSTM) recurrent neural networks (RNNs) exhibit significant performance superiority over FNNs and generalize well to speakers unseen during training~\cite{2014lstm,2015lstm,chenlstm}. Nevertheless, deep RNN-LSTM architectures involve a large number of parameters, meanwhile, the parallelization of computation is impaired due to the inherent sequential nature of RNNs, which excludes its use in many applications. Instead of the frame-wise sequential prediction in RNNs, convolutional neural networks (CNNs) process an input sequence of time frames in parallel. Recently, a one-dimensional (1D) fully convolution network termed the residual temporal convolutional network (ResTCN) employs a deep stack of residual blocks comprised of dilated 1D convolution layers to build a very large receptive field, demonstrating impressive capability in capturing long-range temporal dependencies. ResTCNs have been reported to yield comparable or better results than RNNs on a number of sequence modeling problems, with significantly faster training speed and less parameter overhead~\cite{TCN2018}. ResTCNs have recently been applied to speech enhancement~\cite{GRN,DeepMMSE,tfa} and speaker separation~\cite{zhiqiang2019,convtasnet} with considerable success. More recently, Transformer network architecture, which employs the self-attention mechanism to enable the model to capture long-term dependencies more efficiently, has established state-of-the-art results in numerous speech processing tasks, including speech enhancement~\cite{tgsa,tfaj,conformer-se,dang2022dpt,yu2022dual}.}

\textcolor{black}{To the best of our knowledge, the problem of length generalization has not yet been investigated in Transformer-based speech enhancement. There have been attempts to explore it with Transformers in language modeling \cite{kerple,alibi,anil2022exploring}, machine translation, and image recognition \cite{likhomanenko2021cape}. Anil \textit{et al.} \cite{anil2022exploring} define the length generalization problem and establish that few-shot scratchpad prompting leads to a considerable improvement for pre-trained large language models (LLMs) in length generalization. Press \textit{et al.} \cite{alibi} propose an RPE method termed ALiBi that employs fixed slopes to weight the real relative distance to enable the length generalization for causal language modeling. Later, Chi \textit{et al.} \cite{kerple} propose the kernelized relative position embedding (KERPLE), which demonstrates that the ALiBi is a particular instance of KERPLE and achieves superior generalization performance \cite{kerple} in perplexity. Dubois \textit{et al.} \cite{dubois2020location} present a novel location-based attention to facilitate generalization to longer sequences. Newman \textit{et al.} \cite{newman2020extrapolation} find that generative language models trained to predict the end-of-sequence (EOS) token exhibit worse length generalization ability than those not trained to predict the EOS token. A recent work \cite{yehudai2021local} investigates the capability of graph neural networks (GNNs) to generalize from small to large graphs.}

\section{\textcolor{black}{Positional Encoding}} \label{sec:3}
\textcolor{black}{Existing positional encoding methods in Transformers can be divided into two groups: absolute position encoding (APE) and relative position encoding (RPE).}

\subsection{Absolute Positional Encoding}

\textcolor{black}{APE methods assign a unique position vector or embedding to each time step, which is then summed with the input frame embedding to form the input to the first Transformer layer. APE involves two types of encoding techniques: constant \cite{attention2017} and learnable \cite{bert}.}

\textbf{Sinusoidal Position Embedding}. The original Transformer paper proposes the sinusoidal position embeddings \cite{attention2017} that are fixed and non-learned vectors calculated using the deterministic functions, i.e., cosine and sine functions. Specifically, for the $l$-th time frame, the $d$-th component of position embedding $E_{l,d}$ is given as: 
\begin{equation}
E_{l,d}\!=
\begin{cases}
\sin\left(l \cdot 10000^{-\frac{d}{d_{model}}}\right), & d \text{ is even } \\
\cos\left(l \cdot 10000^{-\frac{d-1}{d_{model}}} \right), & d \text{ is odd }
\end{cases}
\end{equation}
\textcolor{black}{where $d_{model}$ denotes the dimension of the input embedding. It is a commonly used positional encoding scheme in Transformer-based speech enhancement \cite{sepformerstft} and speaker separation \cite{sepformer} as well.}

\textbf{Learnable Absolute Position Embedding}. BERT \cite{bert} replaces the fixed sinusoidal position embeddings with fully-learnable absolute position embeddings $\mathbf{E}\!\in\!\mathbb{R}^{L \times d_\text{model}}$ corresponding to the time frames $l=1,2,...,L$ in an utterance. Here we refer to the learnable APE as ``BERT-Pos''.

\subsection{Relative Positional Encoding}\label{sec:3.2}

\textcolor{black}{APE is restricted to fixed max sequence lengths. Instead of encoding absolute positions, RPE considers relative distances between two elements in a sequence.}

\textbf{Gaussian Bias} \textcolor{black}{(Gauss-Bias) is proposed for acoustic modeling in \cite{gaussbias}, which encodes the relative position through a Gaussian function, where the relative position between $i$-th and $j$-th time frames is encoded as:}
\begin{equation}\label{gauss-bias}
P_{i,j}=\frac{-(i-j)^2}{2 \sigma^2}
\end{equation}
\textcolor{black}{where $\sigma$ denotes a trainable standard deviation parameter. Different parameters are learned for each attention head, and parameters are shared across layers. 
}

\textbf{T5-Bias}. 
\textcolor{black}{In T5 model \cite{T5}, the relative positions $i-j$ are split into a fixed number of buckets with a log-binning strategy and the positions in the same bucket share the same position embedding. The relative position embedding or bias is shared across layers. To be specific, given a bucket of $32$ learnable parameters $\textbf{B}$, for each attention head, the scalar position bias $P_{i,j}$ is extracted from the bucket with the following rule: }
\begin{equation}
P_{i,j}\!=\! 
\begin{cases}
\textbf{B}[\min(15,8\!+\!\lfloor\frac{\log ((|i-j|)/8)}{\log(128/8)} \cdot 8\rfloor)\!+\!16], \quad i\!-\!j \!\leq\! -8 \\
\textbf{B}[|i-j|+16], \qquad \qquad \qquad \qquad \quad \, \!{-8}\!<\! i\!-\!j\!<\!0\\ 
\textbf{B}[i-j], \qquad \qquad \qquad \qquad \qquad \qquad \,\,\, 0\!\leq\!i\!-\!j\!<\!8\\ 
\textbf{B}[\min(15,8\!+\!\lfloor\frac{\log ((i-j) / 8)}{\log (128 / 8)} \cdot 8\rfloor)], \qquad \qquad i\!-\!j\!\geq\! 8
\end{cases}
\end{equation}
\textcolor{black}{Where $\lfloor\cdot\rfloor$ denotes the floor function. We refer to this RPE method as ``T5-Bias''.}

\textbf{Translation-Invariant Self-Attention (TISA)}. \textcolor{black}{The TISA model \cite{tisa} encodes relative positions using radial-basis functions shaped by several trainable parameters, which provides both localized and global attention, defined as:}

\begin{equation}
{P_{i,j}=\sum_{s=1}^S a_s \exp \left(-\left|b_s\right|\left(j-i-c_s\right)^2\right)}
\end{equation}
{where $a_s$, $b_s$, and $c_s$ denote the three trainable parameters included for each kernel $s\!\in\!\left\{1,...,S\right\}$.} Every attention head and layer employs different parameters. We refer to this positional encoding method as the ``TISA''. 

\textbf{DA-Bias}. The distance-aware Transformer \cite{DA} employs a learnable sigmoid function to encode the relative position (referred to as ``DA-Bias''), given as:
\begin{equation}\label{dabias}
    P_{i,j} = \frac{1+\text{exp}(v)}{1+\text{exp}(v-(w|i-j|))}
\end{equation}
where $w$ and $v$ denote two learnable parameters for each head that are used to scale the relative distance and control the upper bound and ascending steepness of the function, respectively. 

\textbf{KERPLE} \cite{kerple} learns the relative position embedding with conditionally positive definite kernels, including the power variant and the logarithmic variant, where the logarithmic variant achieves better performance, given as:
\begin{equation}\label{kerple}
    P_{i,j} = -r_{1}\cdot \text{log}(1+r_{2}|i-j|)
\end{equation}
where $r_{1}>0$ and $r_{2}>0$ are two learnable parameters for each head. Note KERPLE is limited to causal language modeling. Here, we do not constrain our speech enhancement system to be causal.

\textcolor{black}{Among the aforementioned RPE methods}, except for DA-Bias which injects the position information with multiplication operation, the other RPE methods inject position information by adding the relative position embeddings (or bias) with the raw attention scores matrix.

\textcolor{black}{\textbf{Rotary Position Embedding (RoPE)}. The RoPE~\cite{RoPE} injects position information by rotating key and query representations with angles proportional to their absolute positions in each self-attention layer. With this rotation, the attention dot product exhibits the explicit relative position dependency. We refer readers to the original publication~\cite{RoPE} for more details.}

\section{Time-Frequency Neural Speech Enhancement}\label{sec:4}


\subsection{Signal Model}\label{sec:2.1}
\textcolor{black}{Given a clean speech signal $s[n]$, which is corrupted by an uncorrelated additive noise signal $v[n]$, the observed noisy speech signal $x[n]$ can be formulated as:
\begin{equation} \label{equ:1}
x[n] = s[n] + v[n],
\end{equation}
where $n$ denotes the index of discrete-time samples. Applying the short-time Fourier transform (STFT) to the noisy speech signal $x$, the time-frequency representation is obtained:
\begin{equation} \label{equ:2}
X_{l,k} = S_{l,k} + V_{l,k},
\end{equation}
where $X_{l,k}\!\in\!\mathbb{C}$, $S_{l,k}\!\in\!\mathbb{C}$, and $V_{l,k}\!\in\!\mathbb{C}$ respectively represent the complex-valued STFT spectra of the noisy speech $x$, the clean speech $s$, and the noise component $v$,  at the $k\text{-th}$ discrete-frequency bin of the $l\text{-th}$ time frame. In polar form, equation (\ref{equ:2}) can be expressed as:
\begin{equation} \label{equ:3}
|X_{l,k}|e^{j\theta_{X_{l,k}}} = |S_{l,k}|e^{j\theta_{S_{l,k}}} + |V_{l,k}|e^{j\theta_{V_{l,k}}},
\end{equation}
where $|\cdot|$ extracts the spectral magnitude and $\theta$ denotes the spectral phase.}


\subsection{Training Objectives}\label{sec:2.2}
\textcolor{black}{A backbone network model is trained to optimize its training objective for speech enhancement. Studies have shown that one improves the quality and intelligibility of speech in speech enhancement by optimizing the network model in regard to the T-F mask~\cite{tfaj}. \textcolor{black}{Without loss of generality, this study employs \textcolor{blue}{four} commonly used training objectives, which are outlined below.}}




\subsubsection{Magnitude Spectrum}
\textcolor{black}{The magnitude spectrum (MS) of the clean speech $\left|S_{l,k}\right|$ is one typical training objective in spectral mapping methods to speech enhancement. The estimated clean speech MS $|\widehat{S}_{l,k}|$ is employed with the spectral phase of the noisy speech $\theta_{X_{l,k}}$ to obtain the enhanced STFT spectrum, $\widehat{S}_{l,k}\!=\!|\widehat{S}_{l,k}|e^{j\theta_{X_{l,k}}}$. Then, inverse STFT (ISTFT) is applied to reconstruct time-domain waveform of the enhanced speech $\widehat{s}$.}

\subsubsection{Ideal Ratio Mask}

\textcolor{black}{In masking-based methods to speech enhancement, the ideal ratio mask (IRM)~\cite{targets} is one of the most commonly used training objectives, which is given as:
\begin{equation}\label{IRM}
\text{IRM}[l,k]=\left(\frac{|S_{l,k}|^{2}}{|S_{l,k}|^{2}+|V_{l, k}|^{2}}\right)^{\gamma}
\end{equation}
where $\left|V_{l,k}\right|$ denotes the STFT magnitude spectra of the noise and $\gamma$ the scale parameter that is commonly set to 0.5~\cite{overview2018}.}




\subsubsection{Phase-Sensitive Mask}

\textcolor{black}{As an extension of \textcolor{black}{the spectral magnitude mask (SMM)}, the phase-sensitive mask (PSM)~\cite{psm} involves both spectral phase and magnitude error:
\begin{equation}\label{PSM}
\text{PSM}[l,k]=\frac{|S_{l, k}|}{|X_{l, k}|}\cos[\theta_{{S}_{l,k}} -\theta_{{X}_{l,k}}]
\end{equation}
where $\theta_{{S}_{l,k}}$ and $\theta_{{X}_{l,k}}$ denote the spectral phases of the clean speech and the noisy speech, respectively. The introduction of phase error allows the estimated magnitudes to compensate for the use of noisy speech phases and shows a higher signal-to-distortion ratio (SDR) than IRM and SMM~\cite{psm}.}



\textcolor{black}{The value of the IRM lies in the range of 0 and 1. From equation~(\ref{PSM}), it can be found that the PSM value has an upper bound of greater than 1. In this study, we truncate the PSM values to the range of $[0,1]$, to fit the output range of the sigmoidal function. }

\subsubsection{\textcolor{black}{Complex Ideal Ratio Mask}} 

\textcolor{black}{Unlike the aforementioned masks, the complex ideal ratio mask (cIRM)~\cite{cirm} is defined in the complex domain:
\begin{equation}
\text{cIRM}=\frac{X^{r} S^{r} +X^{i} S^{i}}{(X^{r})^2+(X^{i})^2}+j \frac{X^{r} S^{i}-X^{i} S^{r}}{(X^{r})^2+(X^{i})^2}
\end{equation}
where the superscripts $r$ and $i$ denote the real and imaginary components, respectively. The indices $l$ and $k$ are omitted for simplicity. We employ the same compression function in~\cite{cirm} to compress the cIRM.
}


\textcolor{black}{With the real-valued T-F masks, i.e., IRM and PSM, as training objectives, DNNs are optimized to yield masks $\widehat{M}_{l,k}$ at run time. The attained T-F masks are then applied as a suppression rule on the STFT spectral magnitude of the noisy speech $|X_{l,k}|$ to produce a clean version, given as
\begin{equation}
|\widehat{S}_{l,k}|\!=\!|X_{l,k}|\!\cdot\!\widehat{M}_{l,k}.
\end{equation}
Then, the enhanced spectral magnitude $|\widehat{S}_{l,k}|$ is employed with the noisy spectral phase $\theta_{X_{l,k}}$ to recover the waveform of clean speech $\widehat{s}$ through ISTFT operation. \textcolor{black}{The estimated cIRM is multiplied by the complex spectrum of the noisy speech $X$ to compute a clean version.}}

\section{\textcolor{black}{Speech Enhancement With Position-aware Transformer}}
\label{sec:5}

\subsection{\textcolor{black}{Network Architecture}}\label{sec4.1}

\begin{figure}[!t]
\vskip -0.1in
\centering
\begin{subfigure}[t]{0.51\columnwidth}
\centerline{\includegraphics[width=0.99\columnwidth]{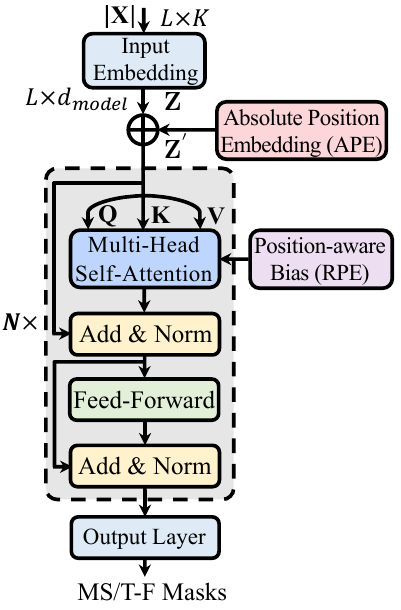}}
\caption{}
\end{subfigure}
\hfill
\begin{subfigure}[t]{0.45\columnwidth}
\centerline{\includegraphics[width=0.99\columnwidth]{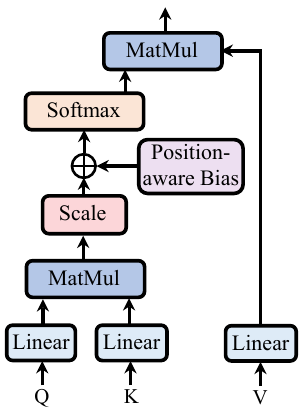}}
\caption{}
\end{subfigure}
\caption{\textcolor{black}{Illustration of (a) the position-aware Transformer backbone network and (b) the position-aware self-attention. $\oplus$ denotes the element-wise summation operation.}}
\vspace{-1.8em}
\label{fig2}
\end{figure}

\textcolor{black}{\textcolor{black}{Fig.~\ref{fig2} (a) illustrates the overall architecture of the position-aware Transformer backbone network for speech enhancement.} The input to the network is the STFT spectral magnitude of the noisy speech, $|\textbf{X}|\!\in\!\mathbb{R}^{L\times K}$, i.e., $L$ time frames of $K$ discrete frequency bins. The embedding layer is a fully-connected (FC) layer that involves a frame-wise layer normalization followed by the ReLU activation function~\cite{tfaj}, which projects the input $|\textbf{X}|$ into a latent representation or embedding with a dimension of $d_{model}$, $\textbf{Z}\!\in\!\mathbb{R}^{L\times d_{model}}$~\cite{tfa}. \textcolor{black}{The APE methods (Sinusoidal PE and BERT-Pos) incorporate the position information by adding the position embedding $\textbf{E}\!\in\!\mathbb{R}^{L\times d_{model}}$ to the input embedding $\textbf{Z}$. The resulting representation $\textbf{Z}^{\prime}\!=\!\textbf{Z}\!+\!\textbf{E}$ is then passed to $N$ stacked Transformer layers for feature transformation. Each Transformer layer consists of two sub-layers. The first one is a multi-head self-attention (MHSA) module, and the second one is a two-layer feed-forward network (FFN). A residual skip connection is employed around each sub-layer. The output layer (on the top of the last Transformer layer) of the network is an FC layer, where a linear layer is used to estimate cIRM, and ReLU and sigmoidal activation functions are applied to output the estimated clean speech magnitude spectrum (MS) and the three real-valued T-F masks (IRM, SMM, and PSM), respectively. Unlike APE which injects position information in the input embedding, RPE includes position information in the self-attention module, which is described in Section~\ref{sec4.2}.
}}

\begin{figure*}[!ht]
\begin{center}
\centerline{\includegraphics[width=1.39\columnwidth]{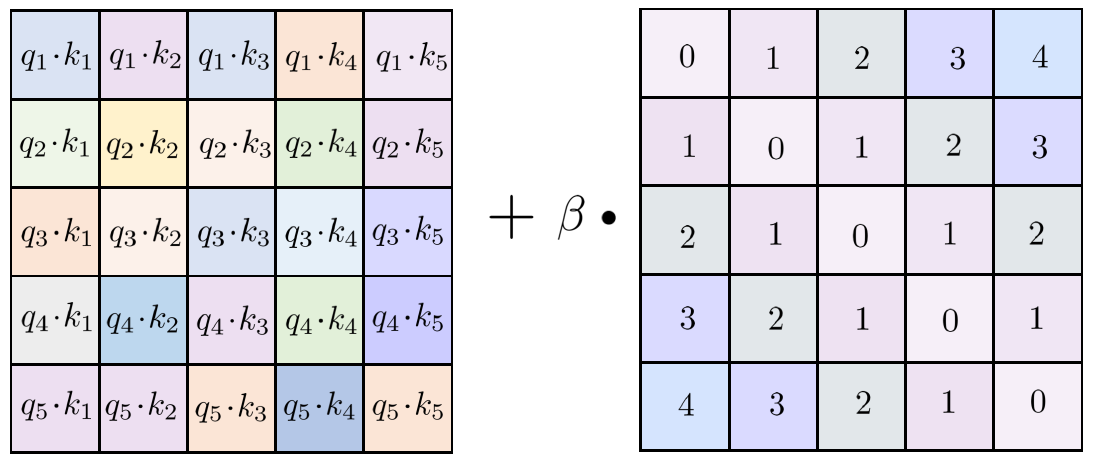}}
\caption{Raw dot-product attention scores are biased by adding a learnable position-aware bias. $\beta$ denotes the learnable head-wise scale parameter, which is different for each attention head shared across Transformer layers.}
\label{LearnLin}
\end{center}
\vspace{-2.0em}
\end{figure*}

\subsection{\textcolor{black}{Position-aware Self-attention}}\label{sec4.2}

\textcolor{black}{Fig.~\ref{fig2} (b) illustrates the workflow of the position-aware self-attention module. It takes as input a set of queries ($\textbf{Q}\!\in \!\mathbb{R}^{L\times d_{model}}$), keys ($\textbf{K}\!\in\!\mathbb{R}^{L\times d_{model}}$), and values ($\textbf{V}\!\in\!\mathbb{R}^{L\times d_{model}}$), and a total of $H$ attention heads are utilized to allow the model to pay attention to different aspects of information, where we denote the head index as $h\!=\!\left\{1,2,3,..,H\right\}$.}


\textcolor{black}{For the $h$-th attention head, $\textbf{Q}$, $\textbf{K}$, and $\textbf{V}$ are first transformed with a linear projection respectively, which leads to $\textbf{Q}^{h}\!=\!\textbf{Q}\textbf{W}_{Q}^{h}$, $\textbf{K}^{h}\!=\!\textbf{K}\textbf{W}_{K}^{h}$, and $\textbf{V}^{h}\!=\!\textbf{V}\textbf{W}_{V}^{h}$, with dimensions of $d_{k}$, $d_{k}$, and $d_{v}$, respectively, where $\{\textbf{W}^{h}_{Q}, \textbf{W}_{K}^{h}\}\!\in\!\mathbb{R}^{d_{model}\times d_{k}}$ and $\textbf{W}_{V}^{h}\in\!\mathbb{R}^{d_{model}\times d_{v}}$ denote learned projection matrices. The scaled dot-product attention is used to compute the attention scores for each attention head. \textcolor{black}{This study explores \textcolor{black}{six} advanced RPE methods (described in Section~\ref{sec:3.2}) to enable the model to generalize from short speech utterances to longer ones. RoPE rotates query $\textbf{Q}$ and key $\textbf{K}$ representations with angles proportional to their absolute positions to incorporate position information in self-attention~\cite{RoPE}. DA-Bias injects position information by multiplying the clipped attention scores with the position-aware bias (refer to Equation (6) in~\cite{DA}).} In the other four RPE methods (Gauss-Bias, T5-Bias, TISA, and KERPLE), the position-aware bias or embeddings are injected with the element-wise summation operation before Softmax normalization, given as:}
\begin{equation}
    \textbf{A}^{h} \left(\textbf{Q}^{h}, \textbf{K}^{h}, \textbf{V}^{h}\right) = \text{softmax}\left(\frac{\textbf{Q}^{h}\textbf{K}^{h\top}}{\sqrt{d_{k}}}+\textbf{P}^{h}\right)\textbf{V}^{h},
\end{equation}
where $\top$ denotes the transpose operation, the head dimension $d_{k}\!=\!d_{v}\!=\!d_{model}/{H}$, and $\textbf{P}^{h}\!\in\!\mathbb{R}^{L\times L}$ the position-aware bias. We refer the reader to the original study \cite{attention2017} for more detailed descriptions of attention scores computation.


Here, we also explore a simpler RPE scheme for length generalization. Similarly, $\textbf{P}^{h}$ is calculated by explicitly using the relative position between time frames. 
To be specific, let us denote the relative distance matrix as $\textbf{R}\!\in\!\mathbb{R}^{L\times L}$, where the $(i,j)$ entry $R_{i,j}\!=\!|i-j|$ denotes the real relative distance between the $i$-th and $j$-th time frames. We attempt to discard complex encoding designs (or formulations) in prior methods (TISA, KERPLE, and DA-Bias) \cite{tisa,kerple,DA} and use only one learnable parameter $\beta^{h}$ to scale $\textbf{R}$ in each attention head to compute the position-aware bias:
\begin{equation}
    P^{h}_{i,j} = \beta^{h} \cdot |i-j|.
\end{equation}
The head-wise scale parameter $\beta^{h}$ is shared across Transformer layers, with a different parameter for different attention heads. Hence, there are $H$ learnable parameters for $H$ attention heads. Unlike Gauss-Bias (Equation (\ref{gauss-bias})), ALiBi, and KERPLE (Equation (\ref{kerple})) that constrain the bias inversely proportional to $R_{i,j}\!=\!|i-j|$, we do not apply such a constraint to learn the position-aware bias. \textcolor{black}{A different parameter $\beta^{h}$ is learned in each attention head to enable different heads to model their diverse preferences on long-term or short-term contextual dependencies with different intensities, allowing the model to capture contextual dependencies more effectively. To be specific, a positive $\beta^{h}$ encourages the attention head to model long-term dependencies, as the attention scores are amplified more intensively by a more positive $P^{h}$. A negative $\beta^{h}$ encourages it to model short-term dependencies, as the attention scores are diminished more intensively by a more negative $P^{h}$.} Since the bias $P^{h}_{i,j}$ is learnable and linear to the real relative position, we refer to it as ``LearnLin''. We can also observe that LearnLin equals the term $w|i-j|$ in DA-Bias (Equation (\ref{dabias})). This study shows that the learnable sigmoidal function is redundant and leads to a performance decrease. Fig.~\ref{LearnLin} provides an example of how the LearnLin bias is injected into the self-attention mechanism (with a sequence length of $L=5$ time frames).

The outputs from all $H$ attention heads are then concatenated and transformed again with a linear projection $\textbf{W}^{O}\!\in\!\mathbb{R}^{d_{model}\times d_{model}}$, forming the MHSA module output:
\begin{equation}
    \text{MHSA}\left(\textbf{Q}, \textbf{K}, \textbf{V}\right) = \text{Concat}[\textbf{A}^{1}\!,\textbf{A}^{2}\!,...,\textbf{A}^{H}]\textbf{W}^{O}.
\end{equation}
The output of the MHSA sub-layer denoted by $\textbf{Y}$ is then passed to the two-layer FFN for two linear transformations, where a ReLU function is applied for the first layer, given as:
\begin{equation}
    \text{FFN}(\textbf{Y}) = \text{ReLU}(\textbf{Y}\textbf{W}_{1} + \textbf{b}_{1})\textbf{W}_{2} + \textbf{b}_{2},
\end{equation}
where $\textbf{W}_{1}\in \mathbb{R}^{d_{model}\times d_{f\!f}}$, $\textbf{W}_{2}\in \mathbb{R}^{d_{f\!f}\times d_{model}}$, $\textbf{b}_{1}\in \mathbb{R}^{d_{f\!f}}$, and $\textbf{b}_{2}\in \mathbb{R}^{d_{model}}$ are projection and bias parameters. The input and output of the FFN sub-layer have a dimension of $d_{model}$, and the dimension of the inner-layer is $d_{f\!f}$.

\section{Experimental Setup}\label{sec:6}

\subsection{Dataset}\label{sec:4.1}
\textcolor{black}{In this section, we provide a detailed description of the clean speech and noise data used in our experiments. For the clean speech data in the training set, we employ the \textit{train-clean-100} subset of the LibriSpeech corpus \cite{Librispeech}, comprising $28\,539$ clean speech utterances from $251$ speakers ($125$ females and $126$ males), with a total duration of approximately 100 hours. The noise data used for training are collected from the Nonspeech dataset \cite{Nonspeech}, the RSG-10 dataset \cite{RSG}, the Environmental Background Noise dataset \cite{ENV1,ENV2}, the Urban Sound dataset \cite{Urban}, the noise subset of the MUSAN corpus \cite{MUSAN}, the QUT-NOISE dataset \cite{QUT}, and the colored noise data (with an $\alpha$ value ranging from $-2$ to $2$ in increments of 0.25) \cite{DeepMMSE}. For testing, we select four noise sources \cite{tfaj}: \textit{factory welding}, \textit{F16}, and \textit{voice babble} from the RSG-10 dataset \cite{RSG} and the \textit{street music} recording no $26\,270$ from the Urban Sound dataset \cite{Urban}. The noise recordings in the dataset that are longer than 30 seconds are split into clips of 30 seconds or less, resulting in a noise set with a total of $6\,809$ noise clips. To perform validation experiments, we randomly selected $1\,000$ clean speech utterances (over 2 seconds) and noise clips from the clean speech and noise data to create a validation set of $1\,000$ noisy speech utterances. The noisy utterances are synthesized by mixing each clean speech with a random segment of one noise clip at a random SNR sampled from the set $\left\{r|r\in \mathbb{Z}: -10 \leq r \leq 20 \right\}$ (dB). As a result, the training set includes a total of $27\,539$ clean speech utterances and $5\,809$ noise clips.}

For evaluation experiments, clean speech utterances are taken from the \textit{test-clean-100} subset of LibriSpeech corpus \cite{Librispeech}. Noise clips are from the aforementioned four real-world noise recordings drawn from the RSG-10 noise dataset \cite{RSG} and the Urban Sound dataset \cite{Urban}, containing two colored (\textit{factory welding} and \textit{F16}) and two non-stationary noise sources (\textit{street music} and \textit{voice babble}).

\begin{figure}[!t]
\centering
\centerline{\includegraphics[width=0.82\columnwidth]{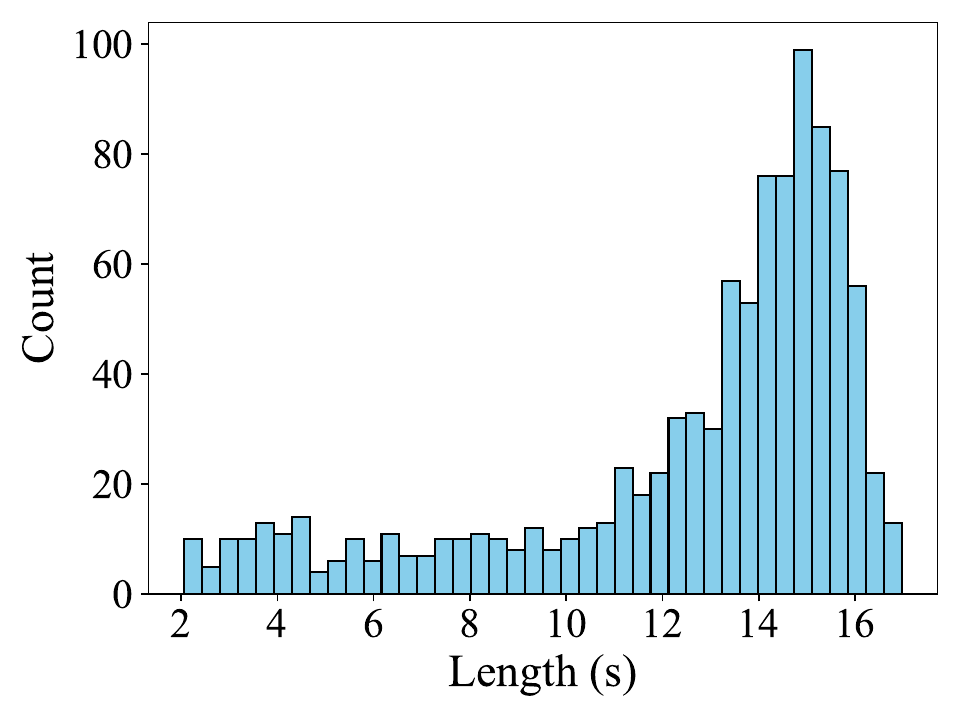}}
\caption{The length distribution over the $1000$ speech utterances in the validation dataset.}
\label{val_length}
\vspace{-1.5em}
\end{figure}

\textbf{Feature Extraction.} \textcolor{black}{All audio signals used in this study are monaural and sampled at a rate of 16 kHz. The noisy speech utterance is segmented into a set of time frames using a square-root-Hann window of length 32 ms ($32\!\times\!16\!=\!512$ time samples), with a hop length of 16 ms ($16\!\times\!16\!=\!256$ time samples). A 512-point fast Fourier transform (FFT) is then applied to each frame, resulting in a 257-dimensional STFT magnitude spectrum as the input, which contains both the DC and Nyquist frequency components.}

\textbf{Length Generalization}. To investigate the length generalization performance of the models, clean speech utterances in the training set are split into 1-second (1s) and 2-second (2s) speech clips for training, respectively. The training details are described in Section \ref{sec:4.2}. For testing, we create six test sets, of which each consists of 400 noisy speech recordings of 1s, 2s, 5s, 10s, 15s, and 20s in length, respectively. The noisy mixtures are generated as follows: we randomly select twenty clean speech utterances (over 1s, 2s, 5s, 10s, 15s, and 20s in length, respectively) from the \textit{test-clean-100} subset for each of the four test noise recordings. Then, a random clip (1s, 2s, 5s, 10s, 15s, and 20s in length) from each clean speech utterance is corrupted by a random clip with the same length cut from the noise recording at five SNR levels, i.e., -5 dB, 0 dB, 5 dB, 10 dB, and 15 dB. In Fig. \ref{val_length}, we provide the distribution of lengths over the $1\ 000$ speech utterances in the aforementioned validation set. We can find that the validation lengths show a broad range (diversity) and most of the speech utterances are more than four times (4s or 8s) the length of the training set (1s or 2s).

\subsection{Implementation Details}\label{sec:4.2}
In this section, we describe the details of model implementations. To evaluate the length generalization of the LearnLin, we employ the standard Transformer backbone network without position embedding (No-Pos) as the baseline \cite{mhanet}. The Transformer model employs $N\!=\!4$ Transformer layers \cite{ripple} and adopt the parameter settings as in \cite{tfaj}: $H\!=\!8$, $d_{model}\!=\!256$, and $d_{f\!f}\!=\!1024$. This study explores several existing APE and RPE methods, which include Sinusoidal \cite{attention2017}, BERT-Pos \cite{bert}, Gauss-Bias \cite{gaussbias}, T5-Bias \cite{T5}, TIAS \cite{tisa}, DA-Bias \cite{DA}, and KERPLE \cite{kerple}, to enable the Transformer speech enhancement model to perform length generalization. For TISA, the number of kernels $S$ is set to 5 \cite{tisa}. It should be noted that KERPLE is initially proposed for causal language modeling. In this study, we derive a non-causal version of KERPLE to study length generalization. Table \ref{params} lists the numbers of trainable parameters for different positional encoding methods.


\textcolor{black}{\textbf{Training Methodology}. Every ten clean speech utterances are selected from the training set and split into clean speech clips of length 1s and 2s (the last clip is dropped), respectively, to create one mini-batch for one training iteration. The noisy mixtures are generated on fly during training. Specifically, each clean speech clip from the mini-batch is corrupted by a random segment of a random noise clip with a randomly sampled SNR from $\left\{r|r\in \mathbb{Z}: -10 \leq r \leq 20 \right\}$ (dB). For each training epoch, the order of picking up clean speech is shuffled randomly. 
For all of the training objectives (i.e., MS, IRM, PSM, \textcolor{black}{and cIRM}), the mean-square error (MSE) is used as the training objective function. Mask approximation is used to learn the T-F mask. For MS, the MSE loss is computed on the power-law compressed magnitude spectrum \cite{2021interactive}. For optimization of the models, we adopt the Adam gradient optimizer with the hyper-parameter settings \cite{attention2017}, $\beta_{1}\!=\!0.9$, $\beta_{2}\!=\!0.98$, and $\epsilon\!=\!1\times10^{-9}$. The gradient clipping is applied to keep the gradient values in the range from -1 to 1. Since the gradient optimization of the self-attention based Transformer models requires a carefully designed learning rate scheduler, a warm-up scheduler is used to dynamically adjust the learning rate \cite{mhanet,attention2017}. Specifically, the formula used to adjust the learning rate is given as \cite{tfaj}:
\begin{equation}
\vspace{-0.2em}
    lr = d_{model}^{-0.5}\cdot \textrm{min} \left(n\_step \cdot w\_steps^{-1.5}, n\_step^{-0.5}\right)
\end{equation}
where $w\_steps$ and $n\_step$ denote the number of warm-up training steps and training steps, respectively. Following~\cite{mhanet}, we use $w\_steps\!=\!40\,000$ for warm-up training scheduler. Our experiments were conducted on an NVIDIA Tesla P100-PCIe-16GB graphics processing unit (GPU).}



\subsection{Assessment Metrics}
\textcolor{black}{Five commonly used objective assessment metrics, which are the perceptual evaluation of speech quality (PESQ) \cite{PESQ}, the extended short-time objective intelligibility (ESTOI) \cite{estoi}, and three composite metrics \cite{composite}, are adopted to comprehensively evaluate enhanced speech signals. Reading all the five speech assessment metrics, we have a higher score to indicate better performance. The PESQ and ESTOI are to evaluate the quality and intelligibility of enhanced speech, respectively. The PESQ score is in the range of $-0.5$ to $4.5$, and the ESTOI score typically ranges from 0 to 1. The three composite metrics describe the mean opinion score (MOS) of the overall speech quality (COVL) \cite{composite}, the signal distortion (CSIG) \cite{composite}, and the background-noise distortion (CBAK) \cite{composite}, respectively. The COVL, CSIG, and CBAK scores range from 0 to 5.}

\input{param.tex}

\begin{figure}[!b]
\centering
\begin{subfigure}[t]{0.5\columnwidth}
\captionsetup{justification=centering}
\centerline{\includegraphics[width=\columnwidth]{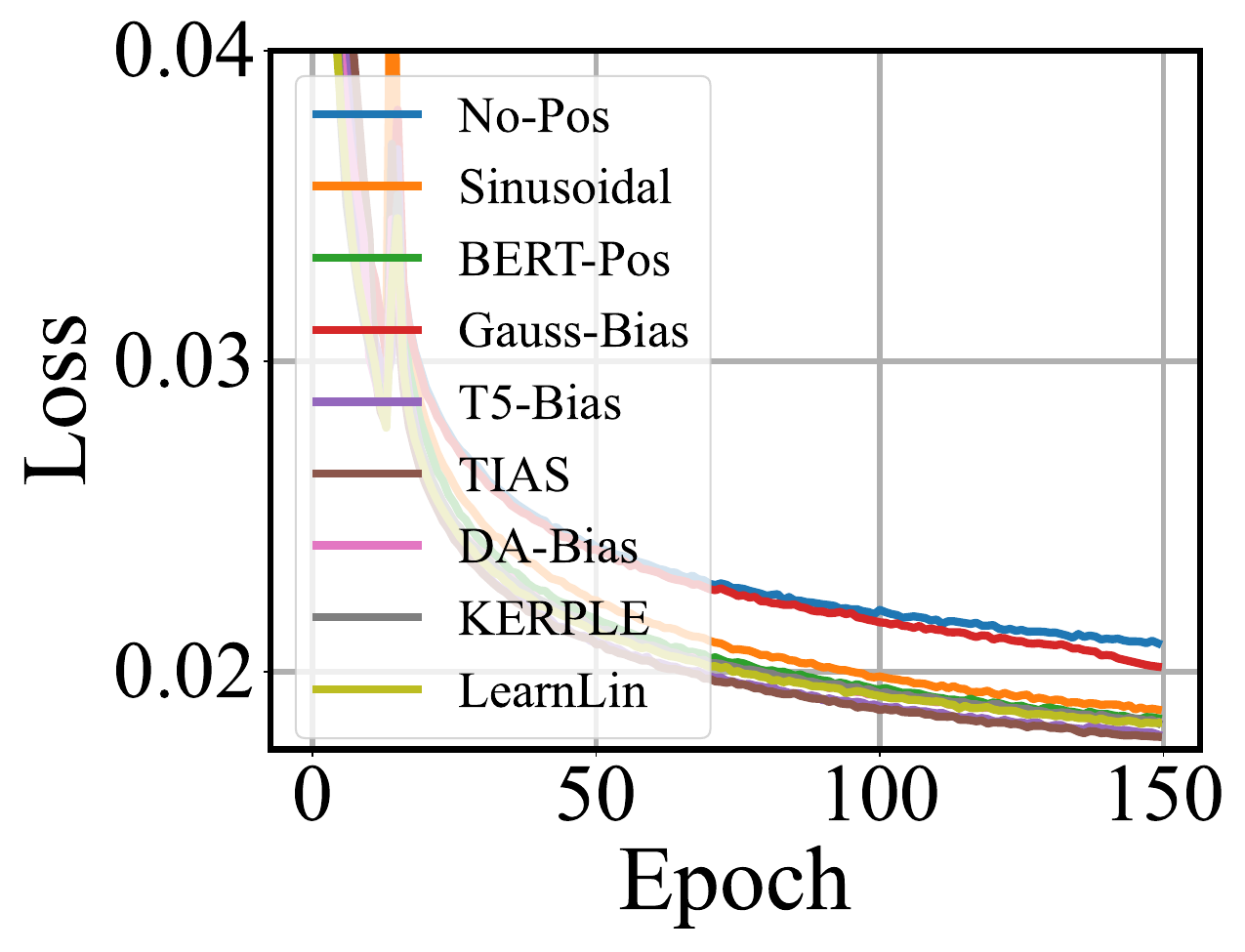}}
\caption{}
\end{subfigure}\hfill
\begin{subfigure}[t]{0.5\columnwidth}
\centerline{\includegraphics[width=\columnwidth]{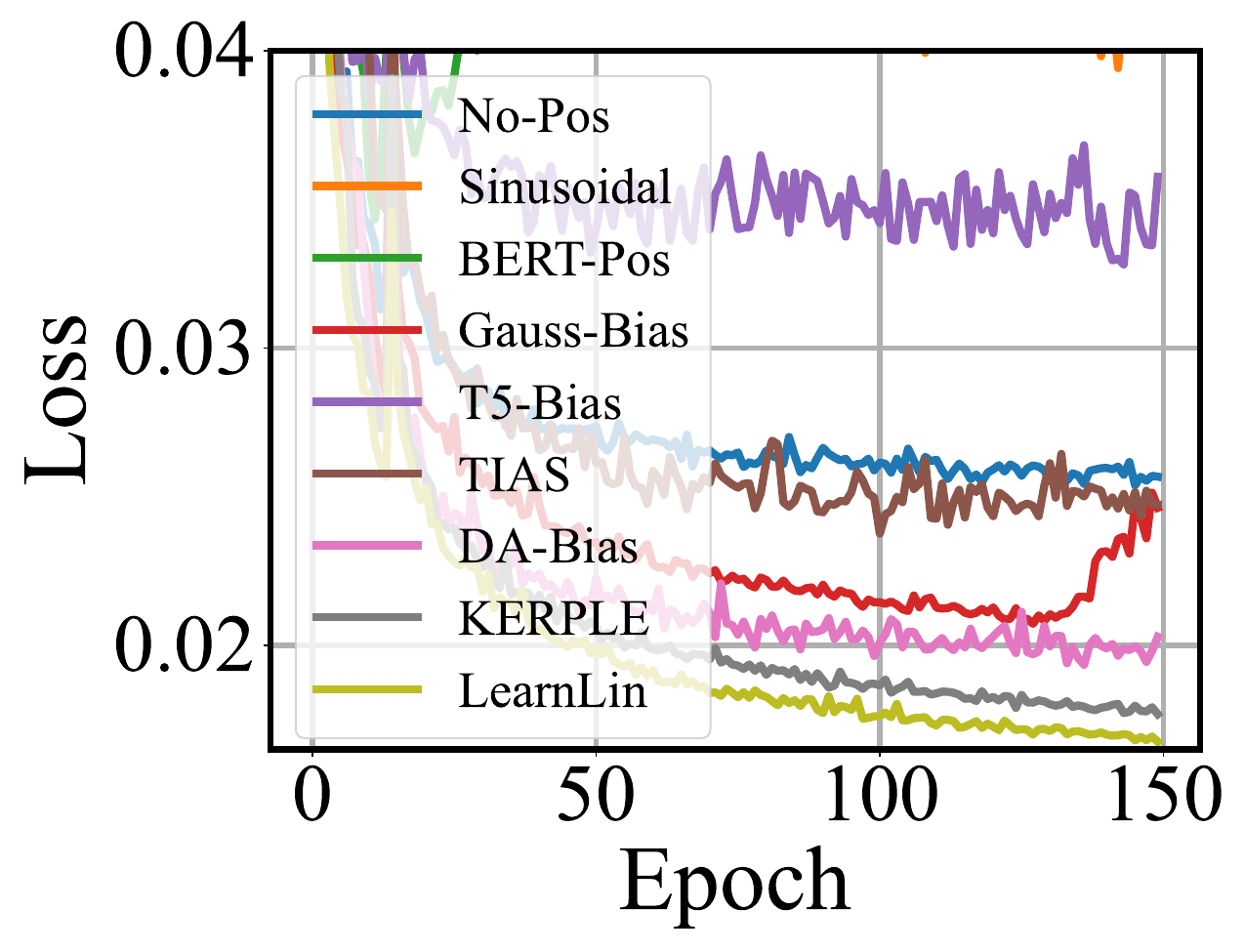}}
\caption{}
\end{subfigure}
\caption{The (a) training loss and (b) validation loss of the models trained using 1s utterances, on MS training objective. }
\vskip -1.0em
\label{train1s_MagMag}
\end{figure}
\section{Experimental Results}\label{sec:7}

\begin{figure}[!t]
\vskip -0.1in
\centering
\begin{subfigure}[t]{0.5\columnwidth}
\captionsetup{justification=centering}
\centerline{\includegraphics[width=\columnwidth]{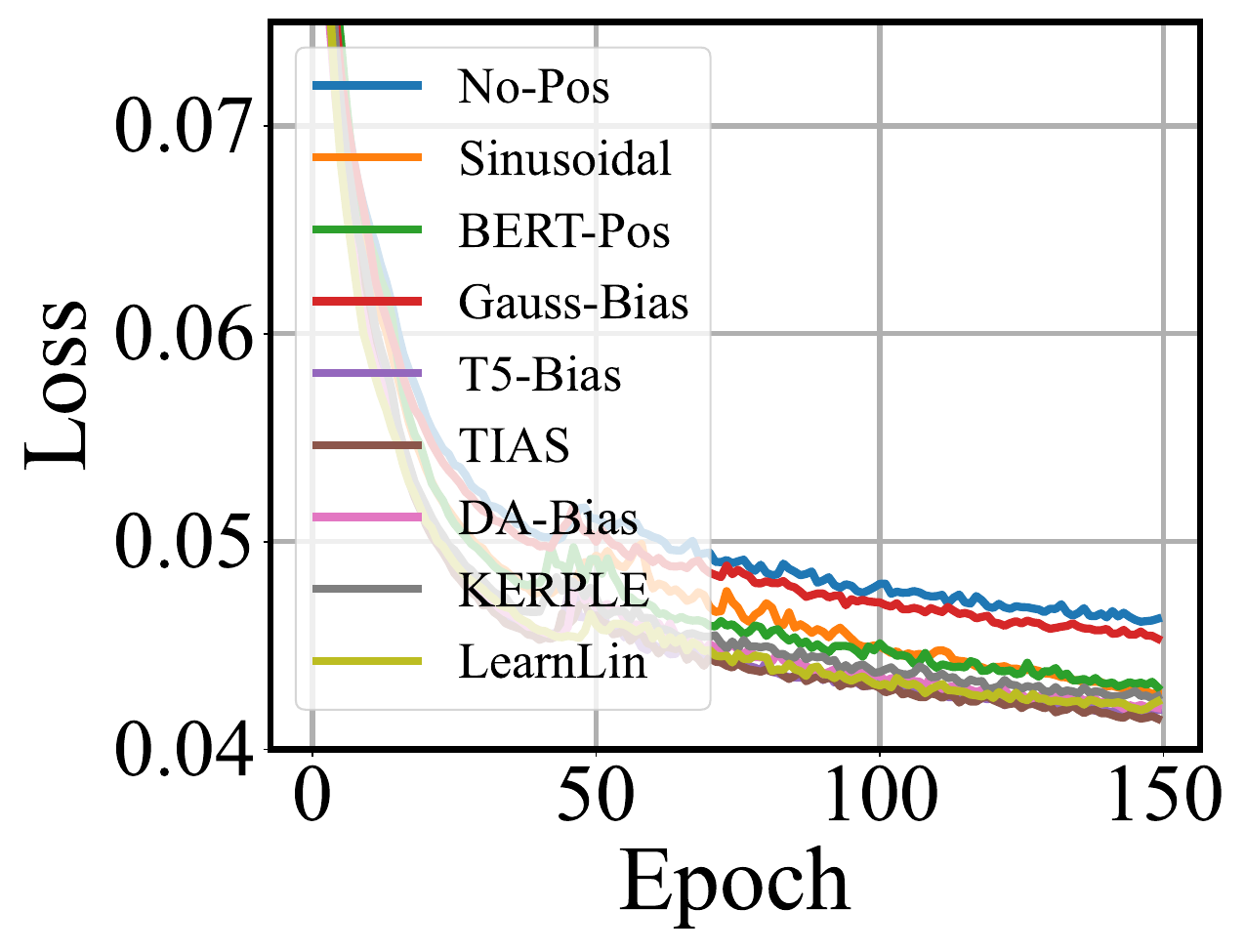}}
\caption{}
\end{subfigure}\hfill
\begin{subfigure}[t]{0.5\columnwidth}
\centerline{\includegraphics[width=\columnwidth]{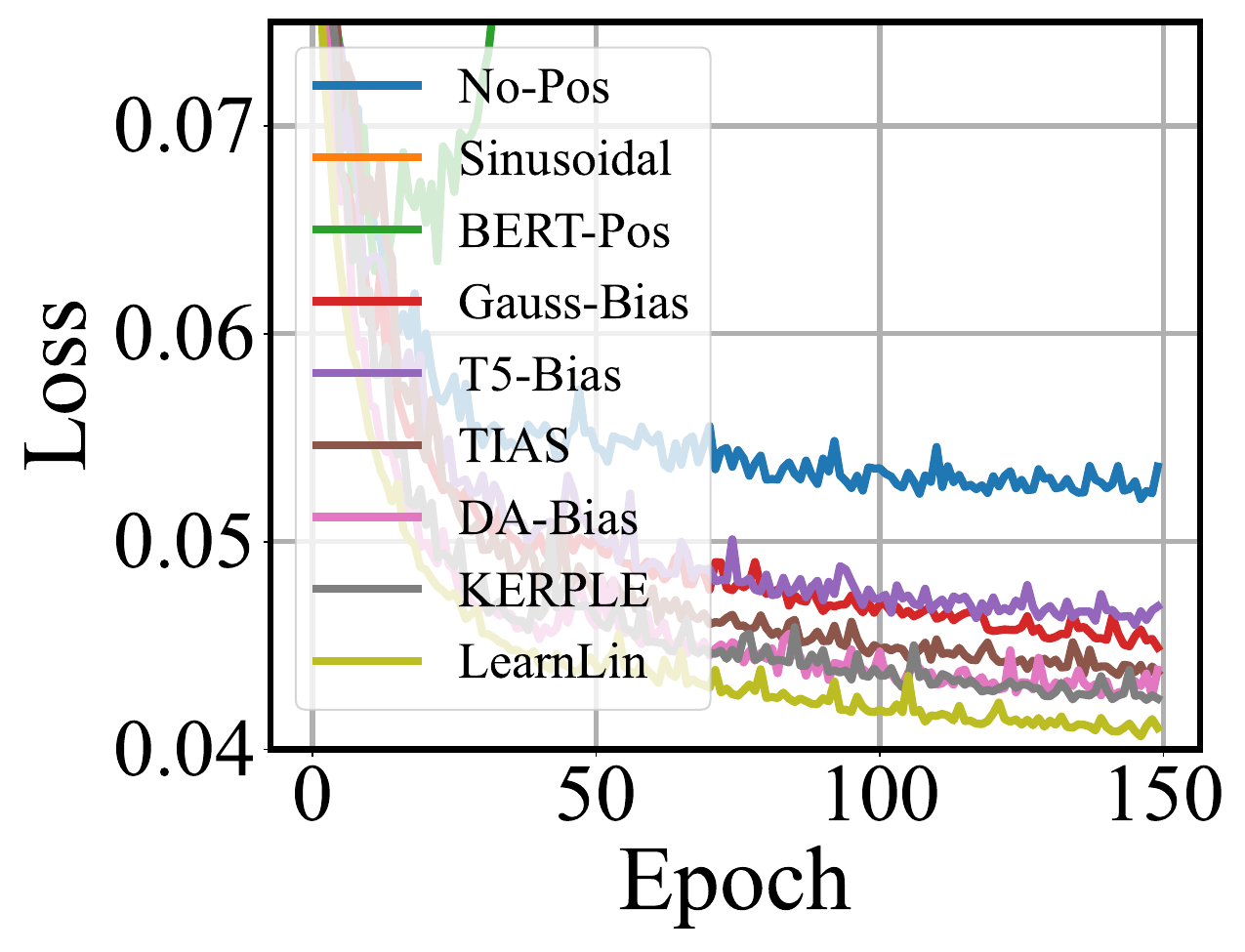}}
\caption{}
\end{subfigure}
\caption{The (a) training loss and (b) validation loss of the models trained using 2s utterances, on IRM training objective. }
\vspace{-0.2em}
\label{train2s_MagIRM}
\end{figure}

\begin{figure}[!t]
\centering
\begin{subfigure}[t]{0.5\columnwidth}
\captionsetup{justification=centering}
\centerline{\includegraphics[width=\columnwidth]{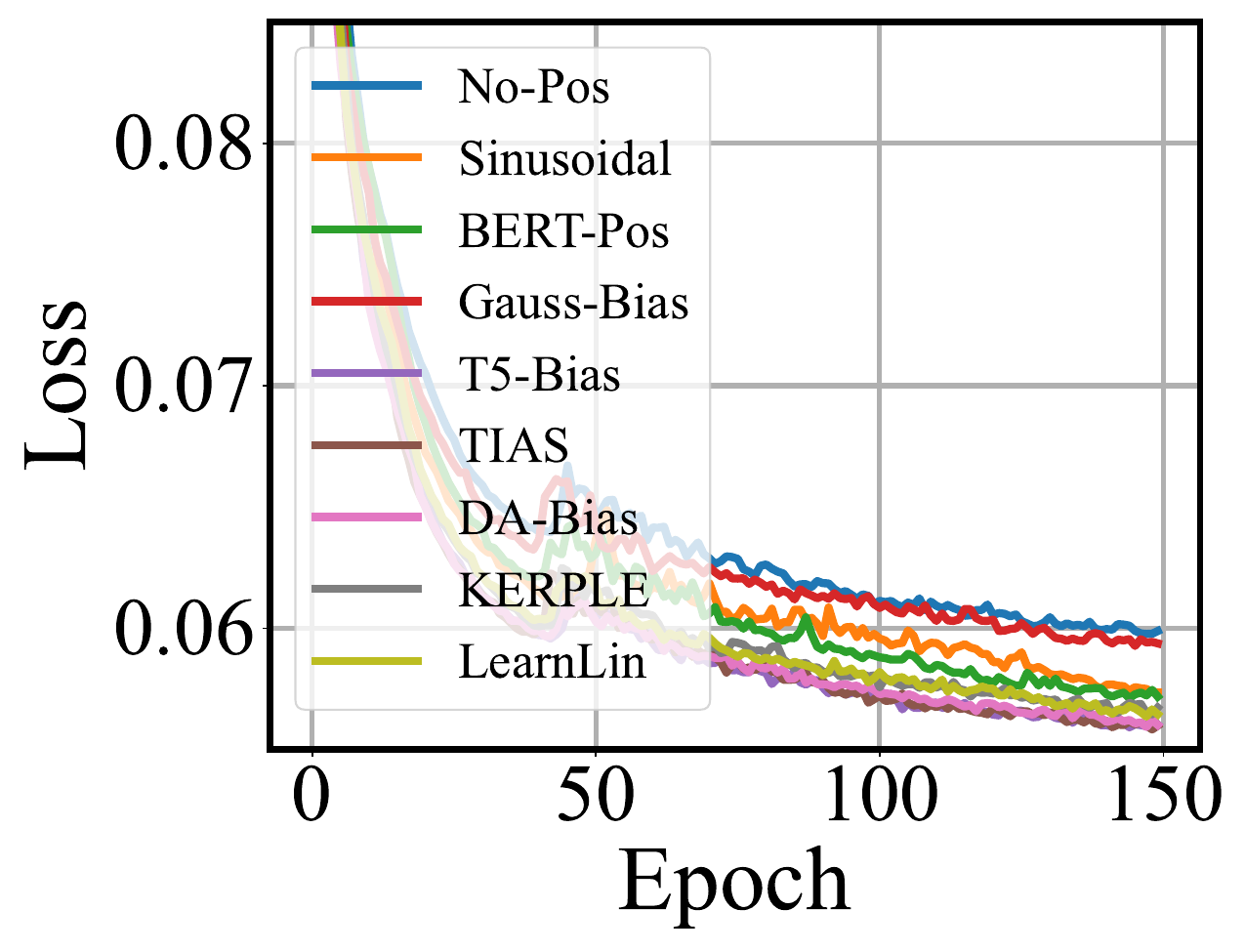}}
\caption{}
\end{subfigure}\hfill
\begin{subfigure}[t]{0.5\columnwidth}
\centerline{\includegraphics[width=\columnwidth]{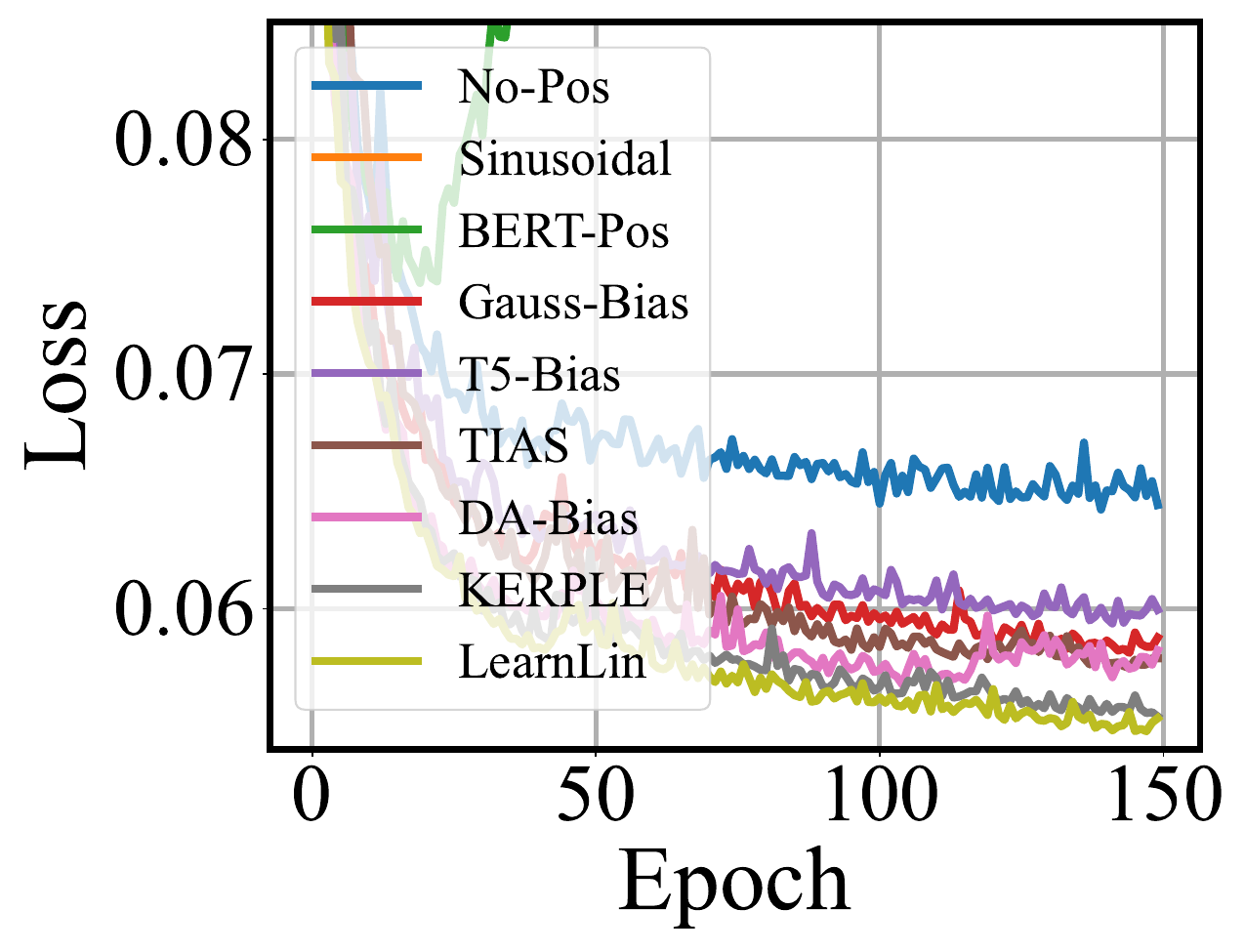}}
\caption{}
\end{subfigure}
\caption{The (a) training loss and (b) validation loss of the models trained using 2s utterances, on PSM training objective. }
\vskip -1.1em
\label{train2s_MagPSM}
\end{figure}

\subsection{Training and Validation Loss}\label{sec:5.1}
In this section, we first observe the training and validation loss values across the models. The loss curves of different models trained with 1s and 2s speech utterances are given in \textcolor{blue}{Fig.~\ref{train1s_MagMag} (MS)} and Fig.~\ref{train2s_MagIRM}-\ref{train2s_MagPSM} (IRM and PSM), respectively, where 150 epochs are used for training models. Similar trends of loss curves are observed on different training objectives for the same training length.

\input{MagMag_1s.tex}

\input{MagIRM_1s.tex}

\input{MagPSM_1s.tex}

\input{MagcIRM_1s.tex}

We observe that the APE methods, i.e., Sinusoidal and BERT-Pos consistently show significant length generalization issues. They yield substantially lower training loss than the model without the positional encoding (No-Pos), while the validation losses are much higher. \textcolor{black}{The comparisons among other methods (such as DA-Bias, KERPLE, and LearnLin) are not obvious when the validation loss curves of Sinusoidal and BERT-Pos are included in the Figures. For ease of comparison, the validation loss curves of Sinusoidal and Bert-Pos are not included in the Figures.} The T5-bias trained using 1s utterances also displays significant length generalization issues. One can observe that RPE methods are more robust to speech length change. The LearnLin provides significantly lower training and validation loss values than No-Pos across training lengths, confirming its excellent length generalization capability. Compared to other strong RPE methods, the LearnLin consistently achieves lower or similar validation loss in a simpler manner, demonstrating its superiority. Among the prior RPE methods, overall, KERPLE shows a better length generalization performance than other methods. 


\subsection{\textcolor{black}{Experiment on Enhancement Performance}}\label{sec:5.2}


In Tables~\ref{tab:magmag1s}--\ref{tab:magcirm1s}, we compare four training objectives, i.e., MS, IRM, PSM, and cIRM, in terms of five metrics. The models are trained using noisy-clean speech pairs of 1s in length, and tested on the noisy mixtures of 1s (without length generalization), 5s, 10s, 15s, and 20s in length, respectively. The numbers denote the average scores over all the noisy test conditions, where the best scores for each metric are highlighted in bold. \textcolor{black}{From Tables~\ref{tab:magmag1s}--\ref{tab:magcirm1s}, it can be clearly observed that the LearnLin substantially improves over the unprocessed noisy speech in terms of five metrics, across all the training objectives and test lengths.} Taking the 1s, 10s, and 20s test lengths as cases, as shown in Table~\ref{tab:magmag1s}, the proposed LearnLin with MS improves PESQ by 0.73, 0.79, and 0.81, ESTOI by 15.43\%, 15.92\%, and 16.50\%, CSIG by 0.89, 0.92 and 0.98, CBAK by 0.66, 0.68 and 0.74, and COVL by 0.84, 0.87 and 0.90, respectively, over the unprocessed noisy recordings. We can observe similar improvements for the proposed model over unprocessed noisy recordings, in IRM, PSM, and cIRM. Comprehensively, PSM yields better performance than the other three among the four training objectives, and the obvious superiority for any one of MS, IRM, and cIRM over the other three is not observed across the models.

In comparison with the model without position embedding (No-Pos), our LearnLin consistently provides significant performance improvements across all test lengths, confirming its excellent length extrapolation property. In the 10s and 20s test length cases, as shown in Table~\ref{tab:magcirm1s}, the LearnLin with cIRM improves PESQ by 0.20 and 0.21, ESTOI by 4.75\% and 5.23\%, CSIG by 0.18 and 0.21, CBAK by 0.12 and 0.15, and COVL by 0.19 and 0.19 respectively, over No-Pos. In the 1s test length case (without length generalization), aside from Gauss-Bias, all the other positional encoding (PE) methods (Sinusoidal, BERT-Pos, T5-Bias, TISA, DA-Bias, KERPLE, and LearnLin) substantially improve No-Pos and attain similar performance scores in five metrics, which is consistent with the loss curves in Fig.~\ref{train1s_MagMag}. It is also obvious that the proposed LearnLin attains the best performance in all five metrics, across four test lengths (5s, 10s, 15s, and 20s) on MS and IRM. On the PSM training objective (shown in Table~\ref{tab:magpsm1s}), aside from that the TISA shows 0.02 CSIG and 0.01 COVL gains over the LearnLin for the 5s test length case, the proposed LearnLin always performs best in all other test cases. On the cIRM training objective, LearnLin outperforms other methods across 10s and 20s test length cases.

\input{MagMag_2s.tex}

\input{MagIRM_2s.tex}
\input{MagPSM_2s.tex}

\input{MagIRM_Causal_1s.tex}

\input{MagIRM_Causal_1s_seg.tex}

\textcolor{black}{Among existing PE methods, for the test length ranging from 5s to 20s, overall KERPLE attains the best length generalization results. Sinusoidal embedding, BERT-Pos, and T5-Bias exhibit lower performance scores compared to the No-Pos model, which demonstrates that they are incapable of length generalization. In particular, sinusoidal embedding and BERT-Pos, which display similar length generalization pathologies, even attain worse ESTOI scores than unprocessed noisy recordings in many cases. DA-Bias, TISA, and Gauss-Bias always achieve better performance than the No-Pos model, confirming their length generalization ability. The overall performance rank in order DA-Bias $>$ TISA $>$ Gauss-Bias, across most cases. \textcolor{black}{We also can observe that with the increase in test length, the performance superiority provided by our LearnLin over these PE methods is more obvious, which significantly demonstrates its capacity and superiority in length generalization. \textcolor{black}{This could be explained by the fact that with the increase in test length, more contextual information becomes available to predict each speech frame and achieve better performance. Moreover, the comparison results (shown in Table~\ref{tab:magirm1s_causeg}) of full-length processing and chunk processing also explain this.}} In the 5s and 20s test length cases, for instance, the proposed LearnLin with IRM provides (shown in Table~\ref{tab:magirm1s}) 0.04 and 0.14 PESQ gains, 1.29\% and 4.46\% ESTOI gains, 0.02 and 0.14 CSIG gains, 0.02 and 0.11 COVL gains, and 0.02 and 0.13 COVL gains respectively, over the DA-Bias.
}

\textcolor{black}{Tables~\ref{tab:magmag2s}--\ref{tab:magpsm1s} report the comparison results among the models that are trained with noisy-clean speech pairs of 2s in length. Similarly, the models are tested on five test lengths, i.e., 2s, 5s, 10s, 15s, and 20s. The boldface numbers represent the highest scores for each metric. From Tables \ref{tab:magmag2s}--\ref{tab:magpsm2s}, we can observe similar performance trends to those shown in Tables \ref{tab:magmag1s}--\ref{tab:magpsm1s}. Again, our proposed LearnLin always outperforms the No-Pos model by a large margin in terms of all five metrics, across different training objectives and all five test lengths, which suggests that our LearnLin enables the model with an excellent capability to extrapolate from shorter utterances to longer utterances.} 

\textcolor{black}{\textcolor{black}{For the 2s test length case (length generalization is not performed), a similar performance trend to that in the 1s test length case is observed, with all the PE methods except Gauss-Bias demonstrating comparable performance in five metrics.} For the test lengths (from 5s to 20s), the LearnLin shows the best performance in almost all the cases but the 5s test length, where TISA with the PSM (shown in Table \ref{tab:magpsm2s}) outperforms the LearnLin by 0.01 in PESQ, CSIG, and COVL. Similarly, among the existing PE methods, KERPLE is more robust than other methods to input length change. Sinusoidal and BERT-Pos exhibit the length generalization pathologies (similar as under 1s training length), where both exhibit significantly lower performance than No-Pos and even yield worse scores than unprocessed speech, especially for long test lengths. Overall, DA-Bias shows a slightly better length generalization property than TISA (on MS, IRM, and PSM), which outperforms T5-Bias and Gauss-Bias. With IRM as the training objective (Table \ref{tab:magirm2s}), TISA improves over T5-Bias by 0.04 and 0.08 in PESQ, 0.53\% and 1.77\% in ESTOI, 0.03 and 0.1 in CSIG, 0.05 and 0.08 in CBAK, and 0.03 and 0.1 in COVL for 5s and 20s test lengths, respectively. Compared to these methods, the superiority of LearnLin is progressively more obvious with the increase in test length. As suggested in Table \ref{tab:magmag2s}, the LearnLin with MS yields 0.01 and 0.09 PESQ gains, 0.52\% and 3.00\% ESTOI gains, 0.02 and 0.13 CSIG gains, 0.01 and 0.06 CBAK gains, and 0.01 and 0.11 COVL gains for 5s and 20s test lengths, respectively, over the DA-Bias. Among T5-bias and Gauss-Bias, T5-Bias achieves better performance for 5s and 10s test lengths, whereas Gauss-Bias shows comparable or better performance for 15s and 20s test lengths.}

\textcolor{black}{In Table~\ref{tab:magirm1s_causal}, we compare the results of the models in causal configuration, with IRM as the training objective and training length of 1s. For simplicity, we report the evaluation results for test lengths of 1s, 10s, and 20s. For the 1s test length case, we can observe that all the RPE schemes achieve similar results. In the test lengths of 10s and 20s cases, aside from that RoPE exhibits lower scores than No-Pos, all the other RPE methods substantially improve over No-Pos in five metrics. Among existing RPE methods, KERPLE provides the best results. LearnLin performs slightly better than KERPLE.}

In Table~\ref{tab:magirm1s_causeg}, we compare the evaluation results of chunk processing and full-length processing. Specifically, we respectively split each 20s-long noisy utterance into $20$ non-overlapped chunks and $39$ overlapped chunks ($50\%$ overlap) of 1s and perform speech enhancement inference separately. Chunk processing mitigates the length generalization problem for No-Pos, RoPE, and T5-Bias. However, chunk processing inevitably suffers from the loss of contextual information due to the boundary effect of small chunks, which leads to obvious performance drops for the models that otherwise benefit from long context. As shown in Table~\ref{tab:magirm1s_causeg}, in comparison to KERPLE and LearnLin, KERPLE-Seg and LearnLin-Seg exhibit decreases of 0.09 and 0.11, 1.76\% and 2.01\%, 0.12 and 0.14, 0.08 and 0.10, and 0.11 and 0.13 in PESQ, ESTOI, CSIG, CBAK, and COVL, respectively. The benefits from overlapped chunk processing are quite limited.

\section{Conclusion}\label{sec:8}
\textcolor{black}{In practical applications, speech enhancement models are often required to perform well on noisy inputs that are longer than the ones used at training time. In this study, we establish and investigate the length generalization problem with Transformer-based speech enhancement models. Several existing relative position embedding (RPE) methods are explored to enable the Transformer speech enhancement model with the ability to learn from shorter speech utterances to generalize longer ones. In addition, this study explores a simpler and more efficient method (LearnLin) to handle length generalization. Extensive speech enhancement experiments on four widely used training objectives (MS, IRM, PSM, and cIRM)} are conducted to evaluate the length generalization capability across the models, with five metrics, i.e., PESQ, ESTOI, CSIG, CBAK, and COVL.

Our experimental results show that the absolute position encoding methods, i.e., Sinusoidal and BERT-Pos are incapable of performing length generalization. In contrast, RPE methods enable the Transformers with the length generalization property. Meanwhile, the comparison results show that our LearnLin achieves better or comparable generalization capability than other RPE methods, i.e., RoPE, Gauss-Bias, T5-Bias, TISA, DA-Bias, and KERPLE, for Transformer-based speech enhancement, with a simpler method. In addition, the comparison results to chunk processing further confirm the superiority of length generalization. In our future works, we will further explore different approaches to length generalization. Also, we plan to explore the LearnLin for length generalization on more speech processing tasks, such as speech separation and speech recognition. 

\ifCLASSOPTIONcaptionsoff
  \newpage
\fi



\bibliographystyle{IEEEtran}
\bibliography{IEEEabrv,./myreference}
\end{document}

%% file: param.tex
\begin{table}[!t]
\centering
    \scriptsize
    \def\arraystretch{1.3}
    \setlength{\tabcolsep}{1.3pt}
    \setlength{\abovetopsep}{0pt}
    \setlength\belowbottomsep{0pt} 
    \setlength\aboverulesep{0pt} 
    \setlength\belowrulesep{0pt}
\caption{The number of trainable parameters for different positional encoding methods, where $L^{'}$ and $S$ (set to 5 as suggested in \cite{tisa}) denote the fixed maximum length (frame numbers) and the kernel numbers in TISA.}
\scalebox{0.97}{\begin{tabular}{l|c|c|c|c|c|c|c|c }
\toprule[1.25pt]
PE & Sinu. & BERT-Pos & Gauss-Bias & TISA & T5-Bias & DA-Bias & KERPLE & LearnLin \\
\hline
\textbf{\#Param.} & -- & $L^{'}\!\cdot\!d_{model}$ & $H$ & $3SHN$ & $32H$ & $2H$ & $2H$ & $H$ \\
\toprule[1.25pt]
\end{tabular}}
\vspace{-1.5em}
\label{params}
\end{table}

%% file: MagMag_1s.tex
\begin{table}[!t]
    \centering
    \scriptsize
    \small
    \def\arraystretch{0.91}
    \setlength{\tabcolsep}{3.3pt}
    \setlength{\abovetopsep}{0pt}
    \setlength\belowbottomsep{0pt} 
    \setlength\aboverulesep{0pt} 
    \setlength\belowrulesep{0pt}
    \caption{Performance evaluation results in terms of PESQ, ESTOI (in \%), CSIG, CBAK, and COVL. All the models are trained (MS as the training objective) with 1s noisy-clean speech pairs and tested on noisy speech utterances of 1s, 5s, 10s, 15s, and 20s in length.}
    \label{tab:magmag1s}
    \begin{tabular}{@{}cl||ccccc@{}}
        \toprule[1.25pt]
         &  & \multicolumn{5}{c}{\textbf{Metrics}}\\  
        Test Len. & Model & PESQ & ESTOI & CSIG & CBAK & COVL \\  
        \hline
        \hline
        \multirow{10}{*}{1s}
        & Noisy    & 1.98 & 54.81 & 2.39 & 1.88 & 1.76 \\
        \cdashline{2-7}
        & No-Pos       & 2.56 & 66.58 & 3.15 & 2.45 & 2.47  \\
        & Sinusoidal   & 2.68 & 70.19 & 3.27 & 2.53 & 2.58  \\
        & BERT-Pos     & 2.69 & 70.15 & 3.26 & 2.53 & 2.58  \\
        & Gauss-Bias   & 2.60 & 67.59 & 3.18 & 2.47 & 2.49  \\
        & T5-Bias      & 2.70 & 70.27 & 3.29 & \textbf{2.55} & \textbf{2.61}  \\
        & TISA         & \textbf{2.71} & \textbf{70.72} & \textbf{3.30} & 2.54 & \textbf{2.61}  \\
        & \textcolor{black}{DA-Bias}           & 2.70 & 70.09 & 3.27 & 2.53 & 2.59  \\
        & \textcolor{black}{KERPLE}       & 2.69 & 69.85 & 3.28 & 2.51 & 2.58  \\
        & LearnLin     & \textbf{2.71} & 70.24 & 3.28 & 2.54 & 2.60 \\
        \hline
        \hline
        \multirow{10}{*}{5s}
        & Noisy    & 1.88 & 54.26 & 2.33 & 1.86 & 1.71 \\
        \cdashline{2-7}
        & No-Pos       & 2.42 & 64.24 & 3.05 & 2.34 & 2.34  \\
        & Sinusoidal   & 2.19 & 57.17 & 2.77 & 2.17 & 2.09  \\
        & BERT-Pos     & 1.96 & 51.86 & 2.60 & 2.02 & 1.92  \\
        & Gauss-Bias   & 2.50 & 65.66 & 3.14 & 2.41 & 2.42  \\
        & T5-Bias      & 2.36 & 62.27 & 2.95 & 2.27 & 2.24  \\
        & TISA         & 2.65 & 68.60 & 3.27 & 2.49 & 2.55  \\
        & \textcolor{black}{DA-Bias}           & 2.67 & 69.92 & 3.28 & 2.53 & 2.58  \\
        & \textcolor{black}{KERPLE}       & 2.67 & 70.40 & 3.29 & 2.50 & 2.57  \\
        & LearnLin     & \textbf{2.70} & \textbf{70.91} & \textbf{3.31} & \textbf{2.53} & \textbf{2.59} \\
        \hline
        \hline
        
        \multirow{10}{*}{10s}
        & Noisy    & 1.92 & 53.19 & 2.40 & 1.87 & 1.75 \\
        \cdashline{2-7}
        & No-Pos       & 2.38 & 61.48 & 2.99 & 2.29 & 2.28  \\
        & Sinusoidal   & 2.11 & 52.17 & 2.67 & 2.09 & 2.01  \\
        & BERT-Pos     & 1.92 & 47.91 & 2.55 & 1.96 & 1.89  \\
        & Gauss-Bias   & 2.51 & 63.68 & 3.12 & 2.39 & 2.41  \\
        & T5-Bias      & 2.27 & 57.54 & 2.82 & 2.17 & 2.12  \\
        & TISA         & 2.56 & 63.18 & 3.08 & 2.38 & 2.39  \\
        & \textcolor{black}{DA-Bias}           & 2.64 & 66.69 & 3.25 & 2.49 & 2.55  \\
        & \textcolor{black}{KERPLE}       & 2.68 & 68.31 & 3.29 & 2.51 & 2.59  \\
        & LearnLin     & \textbf{2.71} & \textbf{69.11} & \textbf{3.32} & \textbf{2.55} & \textbf{2.62}  \\
        \hline
        \hline
        
        \multirow{10}{*}{15s}
        & Noisy        & 1.90 & 52.68 & 2.32 & 1.83 & 1.71 \\
        \cdashline{2-7}
        & No-Pos       & 2.32 & 60.19 & 2.93 & 2.26 & 2.23  \\
        & Sinusoidal    & 2.05 & 50.80 & 2.57 & 2.03 & 1.94  \\
        & BERT-Pos   & 1.97 & 47.03 & 2.47 & 1.93 & 1.85  \\
        & Gauss-Bias   & 2.48 & 63.54 & 3.06 & 2.36 & 2.37  \\
        & T5-Bias      & 2.23 & 58.30 & 2.82 & 2.23 & 2.11  \\
        & TISA         & 2.47 & 60.33 & 2.93 & 2.30 & 2.26  \\
        & \textcolor{black}{DA-Bias}           & 2.58 & 65.65 & 3.18 & 2.45 & 2.49  \\
        & \textcolor{black}{KERPLE}       & 2.64 & 68.39 & 3.23 & 2.47 & 2.53  \\
        & LearnLin     & \textbf{2.68} & \textbf{68.93} & \textbf{3.28} & \textbf{2.49} & \textbf{2.57}  \\
        \hline
        \hline
        
        \multirow{10}{*}{20s}
        & Noisy        & 1.91 & 51.42 & 2.36 & 1.82 & 1.72    \\
        \cdashline{2-7}
        & No-Pos       & 2.34 & 59.07 & 2.98 & 2.28 & 2.26  \\
        & Sinusoidal   & 2.03 & 49.02 & 2.62 & 2.02 & 1.94  \\
        & BERT-Pos     & 1.91 & 45.23 & 2.50 & 1.91 & 1.84 \\
        & Gauss-Bias   & 2.53 & 62.39 & 3.15 & 2.41 & 2.42  \\
        & T5-Bias      & 2.25 & 54.56 & 2.80 & 2.14 & 2.09  \\
        & TISA         & 2.44 & 57.41 & 2.92 & 2.28 & 2.22  \\
        & \textcolor{black}{DA-Bias}      & 2.58 & 63.50 & 3.21 & 2.47 & 2.50  \\
        & \textcolor{black}{KERPLE}       & 2.69 & 67.25 & 3.32 & 2.53 & 2.60  \\
        & LearnLin     & \textbf{2.72} & \textbf{67.92} & \textbf{3.34} & \textbf{2.56} & \textbf{2.62}  \\
        \toprule[1.25pt]
    \end{tabular}
    \vspace{-2.0em}
\end{table}

%% file: MagIRM_1s.tex
\begin{table}[!h]
    \centering
    \scriptsize
    \small
    \def\arraystretch{0.91}
    \setlength{\tabcolsep}{3.3pt}
    \setlength{\abovetopsep}{0pt}
    \setlength\belowbottomsep{0pt} 
    \setlength\aboverulesep{0pt} 
    \setlength\belowrulesep{0pt}
    \caption{Performance evaluation results in terms of PESQ, ESTOI (in \%), CSIG, CBAK, and COVL. All the models are trained (IRM as the training objective) with 1s noisy-clean speech pairs and tested on noisy speech utterances with lengths of 1s, 5s, 10s, 15s, and 20s.}
    \label{tab:magirm1s}
    \begin{tabular}{@{}cl||ccccc@{}}
        \toprule[1.25pt]
         &  & \multicolumn{5}{c}{\textbf{Metrics}}\\  
        Test Len. & Model & PESQ & ESTOI & CSIG & CBAK & COVL\\  
        \hline
        \hline
        \multirow{10}{*}{1s}
        & Noisy    & 1.98 & 54.81 & 2.39 & 1.88 & 1.76 \\
        \cdashline{2-7}
        & No-Pos      & 2.48 & 67.23 & 3.12 & 2.51 & 2.39  \\
        & Sinusoidal   & 2.57 & 70.21 & 3.23 & 2.56 & 2.48  \\
        & BERT-Pos     & 2.58 & 70.61 & 3.22 & 2.58 & 2.49  \\
        & Gauss-Bias   & 2.50 & 67.61 & 3.15 & 2.54 & 2.41  \\
        & T5-Bias      & \textbf{2.60} & \textbf{71.04} & \textbf{3.26} & \textbf{2.59} & \textbf{2.51}  \\
        & TISA         & \textbf{2.60} & 70.86 & \textbf{3.26} & \textbf{2.59} & 2.50  \\
        & \textcolor{black}{DA-Bias}   & 2.58 & 70.65 & 3.25 & 2.57 & 2.50  \\
        & \textcolor{black}{KERPLE}       & 2.57 & 70.13 & 3.24 & 2.57 & 2.49  \\
        & LearnLin     & 2.59 & 71.01 & 3.25 & \textbf{2.59} & 2.50  \\
        \hline
        \hline
        \multirow{10}{*}{5s}
        & Noisy    & 1.88 & 54.26 & 2.33 & 1.86 & 1.71 \\
        \cdashline{2-7}
        & No-Pos       & 2.33 & 64.56 & 2.98 & 2.39 & 2.25  \\
        & Sinusoidal   & 2.17 & 56.22 & 2.73 & 2.24 & 2.04  \\
        & BERT-Pos     & 1.99 & 52.96 & 2.53 & 2.12 & 1.89  \\
        & Gauss-Bias   & 2.42 & 66.43 & 3.09 & 2.48 & 2.34  \\
        & T5-Bias      & 2.31 & 64.06 & 2.97 & 2.37 & 2.23  \\
        & TISA         & 2.53 & 69.87 & 3.22 & 2.54 & 2.46  \\
        & \textcolor{black}{DA-Bias}   & 2.53 & 69.19 & 3.23 & 2.55 & 2.48  \\
        & \textcolor{black}{KERPLE}       & 2.54 & 70.29 & 3.23 & 2.55 & 2.47  \\
        & LearnLin     & \textbf{2.57} & \textbf{70.48} & \textbf{3.25} & \textbf{2.57} & \textbf{2.50} \\

        \hline
        \hline
        
        \multirow{10}{*}{10s}
        & Noisy    & 1.92 & 53.19 & 2.40 & 1.87 & 1.75 \\
        \cdashline{2-7}
        & No-Pos       & 2.32 & 62.48 & 2.99 & 2.35 & 2.24  \\
        & Sinusoidal   & 2.12 & 52.11 & 2.64 & 2.15 & 1.99  \\
        & BERT-Pos     & 1.97 & 50.64 & 2.44 & 2.05 & 1.83  \\
        & Gauss-Bias   & 2.44 & 64.62 & 3.14 & 2.46 & 2.37  \\
        & T5-Bias      & 2.28 & 59.92 & 2.92 & 2.29 & 2.17  \\
        & TISA         & 2.51 & 66.98 & 3.19 & 2.49 & 2.43  \\
        & \textcolor{black}{DA-Bias}   & 2.50 & 66.32 & 3.22 & 2.50 & 2.46  \\
        & \textcolor{black}{KERPLE}       & 2.56 & 68.62 & 3.27 & 2.54 & 2.50  \\
        & LearnLin     & \textbf{2.59} & \textbf{68.89} & \textbf{3.29} & \textbf{2.56} & \textbf{2.53}  \\
        
        \hline
        \hline
        
        \multirow{10}{*}{15s}
        & Noisy        & 1.90 & 52.68 & 2.32 & 1.83 & 1.71 \\
        \cdashline{2-7}
        & No-Pos       & 2.27 & 61.63 & 2.89 & 2.30 & 2.17  \\
        & Sinusoidal   & 2.11 & 51.15 & 2.56 & 2.11 & 1.94  \\
        & BERT-Pos     & 1.92 & 49.68 & 2.39 & 2.01 & 1.79  \\
        & Gauss-Bias   & 2.40 & 64.21 & 3.06 & 2.42 & 2.32  \\
        & T5-Bias      & 2.23 & 58.30 & 2.82 & 2.23 & 2.11  \\
        & TISA         & 2.43 & 65.59 & 3.08 & 2.42 & 2.34  \\
        & \textcolor{black}{DA-Bias}   & 2.44 & 65.05 & 3.12 & 2.44 & 2.37  \\
        & \textcolor{black}{KERPLE}       & 2.52 & 68.31 & 3.20 & 2.49 & 2.45  \\
        & LearnLin     & \textbf{2.55} & \textbf{68.61} & \textbf{3.23} & \textbf{2.53} & \textbf{2.48}  \\

        \hline
        \hline
        
        \multirow{10}{*}{20s}
        & Noisy        & 1.91 & 51.42 & 2.36 & 1.82 & 1.72    \\
        \cdashline{2-7}
        & No-Pos       & 2.29 & 60.33 & 2.95 & 2.32 & 2.21  \\
        & Sinusoidal   & 2.15 & 49.14 & 2.55 & 2.11 & 1.94  \\
        & BERT-Pos     & 1.92 & 48.53 & 2.42 & 1.99 & 1.79  \\
        & Gauss-Bias   & 2.44 & 63.43 & 3.13 & 2.45 & 2.37  \\
        & T5-Bias      & 2.24 & 56.94 & 2.85 & 2.23 & 2.12  \\
        & TISA         & 2.45 & 63.53 & 3.11 & 2.42 & 2.36  \\
        & \textcolor{black}{DA-Bias}   & 2.44 & 63.05 & 3.15 & 2.44 & 2.39  \\
        & \textcolor{black}{KERPLE}       & 2.55 & 67.28 & 3.27 & 2.53 & 2.49  \\
        & LearnLin     & \textbf{2.58} & \textbf{67.51} & \textbf{3.29} & \textbf{2.55} & \textbf{2.52}  \\
        \toprule[1.25pt]
    \end{tabular}
    \vspace{-2.0em}
\end{table}

%% file: MagPSM_1s.tex
\begin{table}[!t]
    \centering
    \scriptsize
    \small
    \def\arraystretch{0.91}
    \setlength{\tabcolsep}{3.3pt}
    \setlength{\abovetopsep}{0pt}
    \setlength\belowbottomsep{0pt} 
    \setlength\aboverulesep{0pt} 
    \setlength\belowrulesep{0pt}
    \caption{Performance evaluation results in terms of PESQ, ESTOI (in \%), CSIG, CBAK, and COVL. All the models are trained (PSM as the training objective) with 1s noisy-clean speech pairs and tested on noisy speech utterances with lengths of 1s, 5s, 10s, 15s, and 20s.}
    \label{tab:magpsm1s}
    \begin{tabular}{@{}cl||ccccc@{}}
        \toprule[1.25pt]
         &  & \multicolumn{5}{c}{\textbf{Metrics}}\\  
        Test Len. & Model & PESQ & ESTOI & CSIG & CBAK & COVL\\  
        \hline
        \hline
        \multirow{10}{*}{1s}
        & Noisy    & 1.88 & 54.26 & 2.33 & 1.86 & 1.71 \\
        \cdashline{2-7}
        & No-Pos       & 2.59 & 68.17 & 3.18 & 2.60 & 2.48  \\
        & Sinusoidal   & 2.70 & 70.96 & 3.29 & 2.68 & 2.59  \\
        & BERT-Pos     & 2.72 & 71.12 & 3.30 & 2.68 & 2.60  \\
        & Gauss-Bias   & 2.60 & 68.19 & 3.20 & 2.61 & 2.49  \\
        & T5-Bias      & \textbf{2.73} & 71.60 & \textbf{3.34} & \textbf{2.70} & \textbf{2.63}  \\
        & TISA         & \textbf{2.73} & \textbf{71.67} & 3.33 & \textbf{2.70} & 2.62  \\
        & \textcolor{black}{DA-Bias}   & 2.71 & 70.91 & 3.29 & 2.68 & 2.58  \\
        & \textcolor{black}{KERPLE}       & 2.71 & 70.51 & 3.29 & 2.69 & 2.60  \\
        & LearnLin     & 2.72 & 70.97 & 3.31 & 2.68 & 2.61 \\
        \hline
        \hline
        \multirow{10}{*}{5s}
        & Noisy    & 1.88 & 54.26 & 2.33 & 1.86 & 1.71 \\
        \cdashline{2-7}
        & No-Pos       & 2.49 & 66.02 & 3.10 & 2.50 & 2.38  \\
        & Sinusoidal   & 2.42 & 62.50 & 2.94 & 2.41 & 2.26  \\
        & BERT-Pos     & 2.11 & 55.49 & 2.61 & 2.21 & 1.96  \\
        & Gauss-Bias   & 2.55 & 67.26 & 3.18 & 2.55 & 2.46  \\
        & T5-Bias      & 2.52 & 65.61 & 3.12 & 2.48 & 2.40  \\
        & TISA         & 2.71 & 70.57 & \textbf{3.36} & 2.66 & \textbf{2.63}  \\
        & \textcolor{black}{DA-Bias}   & 2.70 & 70.11 & 3.34 & 2.67 & 2.61  \\
        & \textcolor{black}{KERPLE}   & 2.69 & 70.27 & 3.32 & 2.64 & 2.59  \\
        & LearnLin     & \textbf{2.72} & \textbf{71.03} & 3.34 & \textbf{2.67} & 2.62 \\
        \hline
        \hline
        
        \multirow{10}{*}{10s}
        & Noisy        & 1.92 & 53.19 & 2.40 & 1.87 & 1.75 \\
        \cdashline{2-7}
        & No-Pos       & 2.46 & 63.89 & 3.09 & 2.45 & 2.36  \\
        & Sinusoidal   & 2.38 & 60.17 & 2.92 & 2.35 & 2.22  \\
        & BERT-Pos     & 2.08 & 53.44 & 2.54 & 2.15 & 1.91  \\
        & Gauss-Bias   & 2.57 & 65.66 & 3.21 & 2.55 & 2.47  \\
        & T5-Bias      & 2.42 & 61.01 & 3.01 & 2.37 & 2.29  \\
        & TISA         & 2.67 & 67.65 & 3.30 & 2.60 & 2.58  \\
        & \textcolor{black}{DA-Bias}   & 2.67 & 67.45 & 3.30 & 2.62 & 2.58  \\
        & \textcolor{black}{KERPLE}   & 2.70 & 68.62 & 3.32 & 2.63 & 2.59  \\
        & LearnLin     & \textbf{2.73}& \textbf{69.49} & \textbf{3.37}& \textbf{2.67}& \textbf{2.64} \\
        \hline
        \hline
        
        \multirow{10}{*}{15s}
        & Noisy        & 1.90 & 52.68 & 2.32 & 1.83 & 1.71 \\
        \cdashline{2-7}
        & No-Pos       & 2.43 & 63.18 & 3.01 & 2.39 & 2.30  \\
        & Sinusoidal   & 2.35 & 58.24 & 2.83 & 2.28 & 2.15  \\
        & BERT-Pos     & 2.05 & 51.42 & 2.46 & 2.09 & 1.85  \\
        & Gauss-Bias   & 2.53 & 65.24 & 3.14 & 2.49 & 2.41  \\
        & T5-Bias      & 2.35 & 58.89 & 2.89 & 2.29 & 2.19  \\
        & TISA         & 2.58 & 65.98 & 3.18 & 2.51 & 2.47  \\
        & \textcolor{black}{DA-Bias}   & 2.59 & 66.18 & 3.21 & 2.54 & 2.49  \\
        & \textcolor{black}{KERPLE}   & 2.67 & 68.24 & 3.25 & 2.58 & 2.54  \\
        & LearnLin     & \textbf{2.71} & \textbf{69.1}5 & \textbf{3.30} & \textbf{2.63} & \textbf{2.59}  \\
        \hline
        \hline
        
        \multirow{10}{*}{20s}
        & Noisy        & 1.91 & 51.42 & 2.36 & 1.82 & 1.72    \\
        \cdashline{2-7}
        & No-Pos       & 2.43 & 61.50 & 3.05 & 2.41 & 2.32  \\
        & Sinusoidal   & 2.39 & 56.71 & 2.87 & 2.29 & 2.19  \\
        & BERT-Pos     & 2.03 & 50.19 & 2.44 & 2.08 & 1.83  \\
        & Gauss-Bias   & 2.55 & 64.11 & 3.19 & 2.52 & 2.45  \\
        & T5-Bias      & 2.40 & 57.02 & 2.88 & 2.28 & 2.18  \\
        & TISA         & 2.60 & 64.07 & 3.20 & 2.51 & 2.48  \\
        & \textcolor{black}{DA-Bias}   & 2.59 & 63.87 & 3.21 & 2.53 & 2.49  \\
        & \textcolor{black}{KERPLE}   & 2.68 & 66.88 & 3.30 & 2.61 & 2.57  \\
        & LearnLin     & \textbf{2.73} & \textbf{67.93} & \textbf{3.36} & \textbf{2.65} & \textbf{2.63}  \\
        \toprule[1.25pt]
    \end{tabular}
    \vspace{-2.0em}
\end{table}

%% file: MagcIRM_1s.tex
\begin{table}[!t]
    \centering
    \scriptsize
    \small
    \def\arraystretch{0.91}
    \setlength{\tabcolsep}{3.3pt}
    \setlength{\abovetopsep}{0pt}
    \setlength\belowbottomsep{0pt} 
    \setlength\aboverulesep{0pt} 
    \setlength\belowrulesep{0pt}
    \caption{\textcolor{black}{Performance evaluation results of PESQ, ESTOI (in \%), CSIG, CBAK, and COVL. All the models are trained (cIRM as the training objective) with 1s noisy-clean speech pairs and tested on noisy speech utterances with lengths of 1s, 10s, and 20s.}}
    \label{tab:magcirm1s}
    \begin{tabular}{@{}cl||ccccc@{}}
        \toprule[1.25pt]
         &  & \multicolumn{5}{c}{\textbf{Metrics}}\\  
        Test Len. & Model & PESQ & ESTOI & CSIG & CBAK & COVL\\  
        \hline
        \hline
        \multirow{7}{*}{1s}
        & Noisy    & 1.98 & 54.81 & 2.39 & 1.88 & 1.76 \\
        \cdashline{2-7}
        & No-Pos       & 2.57 & 67.15 & 2.98 & 2.54 & 2.35  \\
        & T5-Bias      & \textbf{2.69} & \textbf{70.68} & \textbf{3.07} & \textbf{2.62} & \textbf{2.45} \\
        & TISA         & \textbf{2.69} & 70.60 & \textbf{3.07} & \textbf{2.62} & 2.44 \\
        & DA-Bias      & 2.68 & 70.37 & 3.04 & 2.61 & 2.42  \\
        & KERPLE       & 2.65 & 70.27 & 3.02 & 2.60 & 2.41  \\
        & LearnLin     & 2.68 & 70.22 & 3.06 & 2.58 & 2.42  \\
        
        
        \hline
        \hline
        
        \multirow{7}{*}{10s}
        & Noisy    & 1.92 & 53.19 & 2.40 & 1.87 & 1.75 \\
        \cdashline{2-7}
        & No-Pos       & 2.49 & 64.16 & 2.92 & 2.44 & 2.26  \\
        & T5-Bias      & 2.59 & 66.30 & 3.01 & 2.52 & 2.37 \\
        & TISA         & 2.62 & 67.69 & 3.08 & 2.53 & 2.41  \\
        & DA-Bias      & 2.63 & 67.20 & 3.10 & 2.55 & 2.43  \\
        & KERPLE       & 2.67 & 68.09 & \textbf{3.10 }& \textbf{2.56 }& 2.44  \\
        & LearnLin     & \textbf{2.69} & \textbf{68.91} & \textbf{3.10} & \textbf{2.56} & \textbf{2.45}\\

        \hline
        \hline
        
        \multirow{7}{*}{20s}
        & Noisy        & 1.91 & 51.42 & 2.36 & 1.82 & 1.72    \\
        \cdashline{2-7}
        & No-Pos         & 2.49 & 62.13 & 2.90 & 2.40 & 2.24  \\
        & T5-Bias      & 2.54 & 62.32 & 2.96 & 2.43 & 2.31  \\
        & TISA         & 2.59 & 64.75 & 3.08 & 2.47 & 2.38  \\
        & DA-Bias      & 2.61 & 64.17 & 3.08 & 2.49 & 2.40  \\
        & KERPLE       & 2.67 & 66.37 & 3.11 & 2.53 & 2.42  \\
        & LearnLin     & \textbf{2.70 }& \textbf{67.36} & \textbf{3.11 }& \textbf{2.55 }& \textbf{2.43 }\\
        \toprule[1.25pt]
    \end{tabular}
    \vspace{-2.0em}
\end{table}

%% file: MagMag_2s.tex
\begin{table}[!t]
    \centering
    \scriptsize
    \small
    \def\arraystretch{0.91}
    \setlength{\tabcolsep}{3.3pt}
    \setlength{\abovetopsep}{0pt}
    \setlength\belowbottomsep{0pt} 
    \setlength\aboverulesep{0pt} 
    \setlength\belowrulesep{0pt}
    \caption{Performance evaluation results in terms of PESQ, ESTOI (in \%), CSIG, CBAK, and COVL. All the models are trained (MS as the training objective) with 2s noisy-clean speech pairs and tested on noisy speech utterances with lengths of 2s, 5s, 10s, 15s, and 20s.}
    \label{tab:magmag2s}
    
    \begin{tabular}{@{}cl||ccccc@{}}
        \toprule[1.25pt]
         &  & \multicolumn{5}{c}{\textbf{Metrics}}\\  
        Test Len. & Model & PESQ & ESTOI & CSIG & CBAK & COVL\\  
        \hline
        \hline
        \multirow{10}{*}{2s}
        & Noisy        & 1.88 & 53.94 & 2.36 & 1.87 & 1.74 \\
        \cdashline{2-7}
        & No-Pos       & 2.52 & 66.02 & 3.18 & 2.46 & 2.46  \\
        & Sinusoidal   & 2.67 & 69.56 & 3.32 & 2.55 & 2.60  \\
        & BERT-Pos     & 2.69 & 70.22 & 3.34 & 2.57 & 2.63  \\
        & Gauss-Bias   & 2.55 & 66.30 & 3.20 & 2.48 & 2.48  \\
        & T5-Bias      & \textbf{2.72} & \textbf{71.05} & \textbf{3.37} & \textbf{2.59} & \textbf{2.65}  \\
        & TISA         & 2.71 & 70.85 & \textbf{3.37} & 2.58 & \textbf{2.65} \\
        & \textcolor{black}{DA-Bias}           & 2.71 & 70.49 & 3.34 & 2.58 & 2.63  \\
        & \textcolor{black}{KERPLE}   & 2.69 & 70.05 & 3.31 & 2.55 & 2.60  \\
        & LearnLin     & \textbf{2.72} & 70.57 & 3.36 & 2.57 & \textbf{2.65} \\
        \hline
        \hline
        \multirow{10}{*}{5s}
        & Noisy        & 1.88 & 54.26 & 2.33 & 1.86 & 1.71 \\
        \cdashline{2-7}
        & No-Pos       & 2.50 & 66.45 & 3.17 & 2.41 & 2.43  \\
        & Sinusoidal   & 1.48 & 44.94 & 1.95 & 1.81 & 1.48  \\
        & BERT-Pos     & 2.18 & 59.00 & 2.88 & 2.21 & 2.17  \\
        & Gauss-Bias   & 2.56 & 67.49 & 3.22 & 2.45 & 2.49  \\
        & T5-Bias      & 2.70 & 71.04 & 3.38 & 2.54 & 2.65  \\
        & TISA         & 2.71 & 71.14 & 3.39 & 2.56 & 2.67  \\
        & \textcolor{black}{DA-Bias}           & 2.73 & 71.23 & 3.38 & 2.56 & 2.67  \\
        & \textcolor{black}{KERPLE}   & 2.71 & 71.37 & 3.33 & 2.53 & 2.60  \\
        & LearnLin     & \textbf{2.74} & \textbf{71.75} & \textbf{3.40} & \textbf{2.57} & \textbf{2.68} \\
        \hline
        \hline
        
        \multirow{10}{*}{10s}
        & Noisy    & 1.92 & 53.19 & 2.40 & 1.87 & 1.75 \\
        \cdashline{2-7}
        & No-Pos       & 2.47 & 64.05 & 3.15 & 2.37 & 2.42  \\
        & Sinusoidal   & 1.54 & 40.34 & 1.95 & 1.69 & 1.46  \\
        & BERT-Pos     & 1.97 & 50.64 & 2.44 & 2.05 & 1.83  \\
        & Gauss-Bias   & 2.57 & 65.64 & 3.24 & 2.44 & 2.51  \\
        & T5-Bias      & 2.61 & 67.05 & 3.27 & 2.44 & 2.54  \\
        & TISA         & 2.68 & 68.08 & 3.36 & 2.54 & 2.64  \\
        & \textcolor{black}{DA-Bias}           & 2.72 & 68.80 & 3.37 & 2.55 & 2.65  \\
        & \textcolor{black}{KERPLE}   & 2.72 & 69.88 & 3.36 & 2.54 & 2.64  \\
        & LearnLin     & \textbf{2.76} & \textbf{70.22} & \textbf{3.41} & \textbf{2.58} & \textbf{2.69}  \\
        
        \hline
        \hline
        
        \multirow{10}{*}{15s}
        & Noisy        & 1.90 & 52.68 & 2.32 & 1.83 & 1.71 \\
        \cdashline{2-7}
        & No-Pos       & 2.42 & 63.10 & 3.09 & 2.32 & 2.36  \\
        & Sinusoidal   & 1.51 & 37.83 & 1.92 & 1.64 & 1.44  \\
        & BERT-Pos     & 1.97 & 47.03 & 2.47 & 1.93 & 1.86  \\
        & Gauss-Bias   & 2.52 & 65.46 & 3.18 & 2.40 & 2.45  \\
        & T5-Bias      & 2.51 & 64.94 & 3.14 & 2.33 & 2.41  \\
        & TISA         & 2.58 & 66.57 & 3.21 & 2.41 & 2.51  \\
        & \textcolor{black}{DA-Bias}           & 2.66 & 67.67 & 3.31 & 2.50 & 2.60  \\
        & \textcolor{black}{KERPLE}   & 2.69 & 69.59 & 3.33 & 2.50 & 2.61  \\
        & LearnLin     & \textbf{2.72} & \textbf{70.00} & \textbf{3.38} & \textbf{2.54} & \textbf{2.66}  \\
        \hline
        \hline
        
        \multirow{10}{*}{20s}
        & Noisy        & 1.91 & 51.42 & 2.36 & 1.82 & 1.72    \\
        \cdashline{2-7}
        & No-Pos       & 2.28 & 60.33 & 2.95 & 2.31 & 2.21  \\
        & Sinusoidal   & 1.49 & 35.99 & 1.93 & 1.56 & 1.41  \\
        & BERT-Pos     & 1.92 & 48.52 & 2.42 & 1.99 & 1.79  \\
        & Gauss-Bias   & 2.58 & 64.38 & 3.23 & 2.47 & 2.50  \\
        & T5-Bias      & 2.50 & 62.17 & 3.14 & 2.34 & 2.40  \\
        & TISA         & 2.58 & 64.26 & 3.25 & 2.45 & 2.51  \\
        & \textcolor{black}{DA-Bias}           & 2.68 & 65.88 & 3.32 & 2.53 & 2.60  \\
        & \textcolor{black}{KERPLE}   & 2.73 & 68.35 & 3.38 & 2.56 & 2.64  \\
        & LearnLin     & \textbf{2.77} & \textbf{68.88} & \textbf{3.45} & \textbf{2.59} & \textbf{2.71}  \\
        \toprule[1.25pt]
    \end{tabular}
    \vspace{-2.0em}
\end{table}

%% file: MagIRM_2s.tex
\begin{table}[!t]
    \centering
    \scriptsize
    \small
    \def\arraystretch{0.91}
    \setlength{\tabcolsep}{3.3pt}
    \setlength{\abovetopsep}{0pt}
    \setlength\belowbottomsep{0pt} 
    \setlength\aboverulesep{0pt} 
    \setlength\belowrulesep{0pt}
    \caption{Performance evaluation results in terms of PESQ, ESTOI (in \%), CSIG, CBAK, and COVL. All the models are trained (IRM as the training objective) with 2s noisy-clean speech pairs and tested on noisy speech utterances with lengths of 2s, 5s, 10s, 15s, and 20s.}
    \label{tab:magirm2s}
    \begin{tabular}{@{}cl||ccccc@{}}
        \toprule[1.25pt]
         &  & \multicolumn{5}{c}{\textbf{Metrics}}\\  
        Test Len. & Model & PESQ & ESTOI & CSIG & CBAK & COVL\\  
        \hline
        \hline
        \multirow{10}{*}{2s}
        & Noisy        & 1.88 & 53.94 & 2.36 & 1.87 & 1.74 \\
        \cdashline{2-7}
        & No-Pos       & 2.45 & 66.73 & 3.17 & 2.52 & 2.41  \\
        & Sinusoidal   & 2.58 & 70.17 & 3.30 & 2.59 & 2.53  \\
        & BERT-Pos     & 2.58 & 70.09 & 3.29 & 2.59 & 2.53  \\
        & Gauss-Bias   & 2.48 & 67.02 & 3.20 & 2.54 & 2.44  \\
        & T5-Bias      & \textbf{2.60} & 70.70 & 3.32 & 2.61 & 2.55  \\
        & TISA         & 2.59 & \textbf{70.89} & \textbf{3.33} & \textbf{2.63} & \textbf{2.56}  \\
        & \textcolor{black}{DA-Bias}           & 2.59 & 70.27 & 3.30 & 2.62 & 2.54  \\
        & \textcolor{black}{KERPLE}   & 2.57 & 69.69 & 3.29 & 2.60 & 2.53  \\
        & LearnLin     & \textbf{2.60} & 70.26 & 3.32 & 2.62 & \textbf{2.56} \\
        \hline
        \hline
        \multirow{10}{*}{5s}
        & Noisy        & 1.88 & 54.26 & 2.33 & 1.86 & 1.71 \\
        \cdashline{2-7}
        & No-Pos       & 2.42 & 66.64 & 3.12 & 2.48 & 2.36  \\
        & Sinusoidal   & 2.03 & 54.71 & 2.48 & 2.06 & 1.82  \\
        & BERT-Pos     & 1.92 & 53.51 & 2.11 & 1.87 & 1.61  \\
        & Gauss-Bias   & 2.47 & 67.82 & 3.17 & 2.51 & 2.41  \\
        & T5-Bias      & 2.55 & 70.77 & 3.29 & 2.56 & 2.52  \\
        & TISA         & 2.59 & 71.30 & 3.32 & \textbf{2.61} & 2.55  \\
        & \textcolor{black}{DA-Bias}           & 2.58 & 70.72 & 3.30 & 2.58 & 2.53  \\
        & \textcolor{black}{KERPLE}   & 2.57 & 70.76 & 3.29 & 2.59 & 2.52  \\
        & LearnLin     & \textbf{2.61} & \textbf{71.49} & \textbf{3.34} & \textbf{2.61} & \textbf{2.57} \\
        \hline
        \hline
        
        \multirow{10}{*}{10s}
        & Noisy    & 1.92 & 53.19 & 2.40 & 1.87 & 1.75 \\
        \cdashline{2-7}
        & No-Pos       & 2.40 & 64.39 & 3.12 & 2.44 & 2.35  \\
        & Sinusoidal   & 1.95 & 50.74 & 2.29 & 1.89 & 1.69  \\
        & BERT-Pos     & 1.98 & 51.27 & 2.17 & 1.83 & 1.64  \\
        & Gauss-Bias   & 2.49 & 66.23 & 3.22 & 2.51 & 2.45  \\
        & T5-Bias      & 2.51 & 67.53 & 3.26 & 2.51 & 2.48  \\
        & TISA         & 2.58 & 69.02 & 3.34 & 2.58 & 2.56  \\
        & \textcolor{black}{DA-Bias}           & 2.57 & 68.86 & 3.32 & 2.57 & 2.54  \\
        & \textcolor{black}{KERPLE}   & 2.59 & 69.00 & 3.32 & 2.59 & 2.55  \\
        & LearnLin     & \textbf{2.62} & \textbf{69.99} & \textbf{3.36} & \textbf{2.60} & \textbf{2.58}  \\
        
        \hline
        \hline
        
        \multirow{10}{*}{15s}
        & Noisy        & 1.90 & 52.68 & 2.32 & 1.83 & 1.71 \\
        \cdashline{2-7}
        & No-Pos       & 2.35 & 63.34 & 3.03 & 2.38 & 2.28  \\
        & Sinusoidal   & 1.92 & 48.21 & 2.23 & 1.82 & 1.65  \\
        & BERT-Pos     & 1.97 & 50.87 & 2.14 & 1.80 & 1.62  \\
        & Gauss-Bias   & 2.46 & 65.95 & 3.15 & 2.47 & 2.40  \\
        & T5-Bias      & 2.44 & 66.11 & 3.13 & 2.43 & 2.37  \\
        & TISA         & 2.51 & 68.01 & 3.22 & 2.51 & 2.46  \\
        & \textcolor{black}{DA-Bias}           & 2.52 & 67.99 & 3.23 & 2.51 & 2.47  \\
        & \textcolor{black}{KERPLE}   & 2.55 & 68.73 & 3.25 & 2.54 & 2.49  \\
        & LearnLin     & \textbf{2.59} & \textbf{69.65} & \textbf{3.29} & \textbf{2.55} & \textbf{2.52}  \\
        \hline
        \hline
        
        \multirow{10}{*}{20s}
        & Noisy        & 1.91 & 51.42 & 2.36 & 1.82 & 1.72    \\
        \cdashline{2-7}
        & No-Pos       & 2.35 & 61.73 & 3.07 & 2.39 & 2.31  \\
        & Sinusoidal   & 1.90 & 46.25 & 2.17 & 1.80 & 1.62  \\
        & BERT-Pos     & 1.97 & 49.82 & 2.20 & 1.80 & 1.65  \\
        & Gauss. Bias  & 2.49 & 64.69 & 3.21 & 2.49 & 2.43  \\
        & T5 Bias      & 2.44 & 64.23 & 3.15 & 2.42 & 2.38  \\
        & TISA         & 2.52 & 66.00 & 3.25 & 2.50 & 2.48  \\
        & \textcolor{black}{DA-Bias}           & 2.52 & 66.27 & 3.27 & 2.53 & 2.50  \\
        & \textcolor{black}{KERPLE}   & 2.57 & 67.52 & 3.30 & 2.56 & 2.53  \\
        & LearnLin     & \textbf{2.62} & \textbf{68.61} & \textbf{3.36} & \textbf{2.59} & \textbf{2.58}  \\
        \toprule[1.25pt]
    \end{tabular}
    \vspace{-2.0em}
\end{table}

%% file: MagPSM_2s.tex
\begin{table}[!t]
    \centering
    \scriptsize
    \small
    \def\arraystretch{0.91}
    \setlength{\tabcolsep}{3.3pt}
    \setlength{\abovetopsep}{0pt}
    \setlength\belowbottomsep{0pt} 
    \setlength\aboverulesep{0pt} 
    \setlength\belowrulesep{0pt}
    \caption{Performance evaluation results in terms of PESQ, ESTOI (in \%), CSIG, CBAK, and COVL. All the models are trained (PSM as the training objective) with 2s noisy-clean speech pairs and tested on noisy speech utterances with lengths of 2s, 5s, 10s, 15s, and 20s.}
    \label{tab:magpsm2s}
    \begin{tabular}{@{}cl||ccccc@{}}
         \toprule[1.25pt]
         &  & \multicolumn{5}{c}{\textbf{Metrics}}\\  
        Test Len. & Model & PESQ & ESTOI & CSIG & CBAK & COVL\\  
        \hline
        \hline
        \multirow{10}{*}{2s}
        & Noisy    & 1.88 & 53.94 & 2.36 & 1.87 & 1.74 \\
        \cdashline{2-7}
        & No-Pos       & 2.58 & 67.19 & 3.23 & 2.61 & 2.50  \\
        & Sinusoidal   & 2.71 & 70.23 & 3.33 & 2.69 & 2.61  \\
        & BERT-Pos     & 2.70 & 70.55 & 3.33 & 2.70 & 2.62  \\
        & Gauss-Bias   & 2.60 & 67.49 & 3.24 & 2.62 & 2.52  \\
        & T5-Bias      & \textbf{2.74} & 71.15 & 3.37 & 2.71 & \textbf{2.65}  \\
        & TISA         & \textbf{2.74} & \textbf{71.33} & \textbf{3.38} & 2.71 & \textbf{2.65}  \\
        & \textcolor{black}{DA-Bias}   & 2.73 & 70.94 & 3.37 & 2.71 & 2.65  \\
        & \textcolor{black}{KERPLE}    & 2.71 & 70.34 & 3.34 & 2.69 & 2.62  \\
        & LearnLin     & 2.72 & 70.81 & 3.36 & \textbf{2.72} & 2.64  \\
        \hline
        \hline
        \multirow{10}{*}{5s}
        & Noisy    & 1.88 & 54.26 & 2.33 & 1.86 & 1.71 \\
        \cdashline{2-7}
        & No-Pos       & 2.56 & 67.40 & 3.19 & 2.57 & 2.47  \\
        & Sinusoidal   & 1.89 & 53.44 & 2.31 & 1.91 & 1.71  \\
        & BERT-Pos     & 2.06 & 59.48 & 2.61 & 2.10 & 1.92  \\
        & Gauss-Bias   & 2.61 & 68.40 & 3.25 & 2.60 & 2.52  \\
        & T5-Bias      & 2.74 & 71.29 & 3.39 & 2.70 & 2.66  \\
        & TISA         & \textbf{2.76} & 71.80 & \textbf{3.41} & 2.69 & \textbf{2.68}  \\
        & \textcolor{black}{DA-Bias}   & 2.74 & 71.86 & \textbf{3.41} & 2.71 & 2.67  \\
        & \textcolor{black}{KERPLE}    & 2.74 & 71.38 & 3.37 & 2.67 & 2.64  \\
        & LearnLin     & 2.75 & \textbf{71.88} & 3.40 & \textbf{2.71} & 2.67 \\
        \hline
        \hline
        
        \multirow{10}{*}{10s}
        & Noisy    & 1.92 & 53.19 & 2.40 & 1.87 & 1.75 \\
        \cdashline{2-7}
        & No-Pos       & 2.52 & 65.35 & 3.19 & 2.53 & 2.45  \\
        & Sinusoidal   & 1.94 & 53.23 & 2.37 & 1.92 & 1.76  \\
        & BERT-Pos     & 1.99 & 54.68 & 2.47 & 1.97 & 1.82  \\
        & Gauss-Bias   & 2.61 & 66.90 & 3.27 & 2.61 & 2.54  \\
        & T5-Bias      & 2.69 & 67.99 & 3.35 & 2.64 & 2.62  \\
        & TISA         & 2.72 & 69.28 & 3.38 & 2.65 & 2.65  \\
        & \textcolor{black}{DA-Bias}   & 2.73 & 69.54 & 3.41 & 2.68 & 2.67  \\
        & \textcolor{black}{KERPLE}    & 2.74 & 69.95 & 3.39 & 2.67 & 2.65  \\
        & LearnLin     & \textbf{2.75} & \textbf{70.40} & \textbf{3.42} & \textbf{2.72} & \textbf{2.69}  \\
        \hline
        \hline
        
        \multirow{10}{*}{15s}
        & Noisy        & 1.90 & 52.68 & 2.32 & 1.83 & 1.71 \\
        \cdashline{2-7}
        & No-Pos       & 2.47 & 64.44 & 3.10 & 2.46 & 2.37  \\
        & Sinusoidal   & 1.92 & 52.78 & 2.28 & 1.88 & 1.70  \\
        & BERT-Pos     & 1.96 & 52.49 & 2.40 & 1.89 & 1.76  \\
        & Gauss-Bias   & 2.59 & 66.52 & 3.22 & 2.56 & 2.49  \\
        & T5-Bias      & 2.60 & 66.62 & 3.24 & 2.54 & 2.51  \\
        & TISA         & 2.65 & 68.18 & 3.27 & 2.56 & 2.55  \\
        & \textcolor{black}{DA-Bias}   & 2.69 & 68.85 & 3.36 & 2.64 & 2.63  \\
        & \textcolor{black}{KERPLE}    & 2.72 & 69.59 & 3.35 & 2.64 & 2.62  \\
        & LearnLin     & \textbf{2.74} & \textbf{70.18} & \textbf{3.38} & \textbf{2.68} & \textbf{2.65}  \\
        \hline
        \hline
        
        \multirow{10}{*}{20s}
        & Noisy        & 1.91 & 51.42 & 2.36 & 1.82 & 1.72    \\
        \cdashline{2-7}
        & No-Pos       & 2.49 & 62.73 & 3.13 & 2.47 & 2.40  \\
        & Sinusoidal   & 1.91 & 51.37 & 2.33 & 1.86 & 1.72  \\
        & BERT-Pos     & 1.94 & 51.55 & 2.40 & 1.87 & 1.75  \\
        & Gauss-Bias   & 2.62 & 65.50 & 3.27 & 2.59 & 2.53  \\
        & T5-Bias      & 2.61 & 64.20 & 3.26 & 2.54 & 2.52  \\
        & TISA         & 2.66 & 66.12 & 3.31 & 2.58 & 2.58  \\
        & \textcolor{black}{DA-Bias}   & 2.71 & 67.27 & 3.39 & 2.64 & 2.64  \\
        & \textcolor{black}{KERPLE}    & 2.74 & 68.36 & 3.39 & 2.65 & 2.65  \\
        & LearnLin     & \textbf{2.78} & \textbf{68.85} & \textbf{3.43} & \textbf{2.70} & \textbf{2.69}  \\
        \toprule[1.25pt]
    \end{tabular}
    \vspace{-1.5em}
\end{table}

%% file: MagIRM_Causal_1s.tex
\begin{table}[!t]
    \centering
    \scriptsize
    \small
    \def\arraystretch{0.91}
    \setlength{\tabcolsep}{3.3pt}
    \setlength{\abovetopsep}{0pt}
    \setlength\belowbottomsep{0pt} 
    \setlength\aboverulesep{0pt} 
    \setlength\belowrulesep{0pt}
    \caption{\textcolor{black}{Performance evaluation results of PESQ, ESTOI (in \%), CSIG, CBAK, and COVL in causal configuration. All the models are trained (IRM as the training objective) with 1s noisy-clean speech pairs and tested on noisy speech utterances with lengths of 1s, 10s, and 20s.}}
    \label{tab:magirm1s_causal}
    \begin{tabular}{@{}cl||ccccc@{}}
        \toprule[1.25pt]
         &  & \multicolumn{5}{c}{\textbf{Metrics}}\\  
        Test Len. & Model & PESQ & ESTOI & CSIG & CBAK & COVL\\  
        \hline
        \hline
        \multirow{8}{*}{1s}
        & Noisy    & 1.98 & 54.81 & 2.39 & 1.88 & 1.76 \\
        \cdashline{2-7}
        & No-Pos       & 2.43 & 66.32 & 3.06 & 2.45 & 2.33  \\
        & RoPE         & 2.48 & 66.38 & 3.09 & \textbf{2.49 }& 2.37  \\
        & T5-Bias      & 2.48 & 66.68 & \textbf{3.10 }& \textbf{2.49 }& \textbf{2.40} \\
        & TISA         & 2.47 & 66.32 & 3.09 & 2.47 & 2.38  \\
        & \textcolor{black}{DA-Bias}      & \textbf{2.49} & \textbf{66.75} & \textbf{3.10 }& 2.47 & 2.39  \\
        & \textcolor{black}{KERPLE}       & 2.46 & 66.35 & \textbf{3.10 }& 2.48 & 2.37  \\
        & LearnLin     & 2.48 & 66.72 & \textbf{3.10 }& \textbf{2.49 }& 2.38  \\
        
        
        \hline
        \hline
        
        \multirow{8}{*}{10s}
        & Noisy    & 1.92 & 53.19 & 2.40 & 1.87 & 1.75 \\
        \cdashline{2-7}
        & No-Pos       & 2.25 & 60.09 & 2.91 & 2.26 & 2.16  \\
        & RoPE         & 2.16 & 57.57 & 2.81 & 2.19 & 2.07  \\
        & T5-Bias      & 2.38 & 63.86 & 3.08 & 2.41 & 2.32 \\
        & TISA         & 2.44 & 65.36 & 3.15 & 2.45 & 2.38  \\
        & \textcolor{black}{DA-Bias}   & 2.45 & 65.31 & 3.16 & 2.45 & 2.39  \\
        & \textcolor{black}{KERPLE}    & 2.46 & 65.54 & 3.19 & 2.48 & 2.41  \\
        & LearnLin     & \textbf{2.48} & \textbf{65.77} & \textbf{3.20} & \textbf{2.50} & \textbf{2.43} \\
        
        \hline
        \hline
        
        
        
        \multirow{8}{*}{20s}
        & Noisy        & 1.91 & 51.42 & 2.36 & 1.82 & 1.72    \\
        \cdashline{2-7}
        & No-Pos         & 2.15 & 56.56 & 2.80 & 2.15 & 2.05  \\
        & RoPE         & 2.10 & 54.46 & 2.73 & 2.11 & 2.01  \\
        & T5-Bias      & 2.35 & 61.67 & 3.06 & 2.36 & 2.28  \\
        & TISA         & 2.41 & 63.52 & 3.11 & 2.41 & 2.34  \\
        & \textcolor{black}{DA-Bias}   & 2.44 & 63.32 & 3.16 & 2.43 & 2.38  \\
        & \textcolor{black}{KERPLE}       & 2.45 & 64.13 & 3.17 & 2.45 & 2.40  \\
        & LearnLin     & \textbf{2.48} & \textbf{64.59} & \textbf{3.19} & \textbf{2.48} & \textbf{2.43}  \\
        \toprule[1.25pt]
    \end{tabular}
    \vspace{-0.5em}
\end{table}

%% file: MagIRM_Causal_1s_seg.tex
\begin{table}[!t]
    \centering
    \scriptsize
    \small
    \def\arraystretch{0.91}
    \setlength{\tabcolsep}{3.0pt}
    \setlength{\abovetopsep}{0pt}
    \setlength\belowbottomsep{0pt} 
    \setlength\aboverulesep{0pt} 
    \setlength\belowrulesep{0pt}
    \caption{Performance evaluation results of PESQ, ESTOI (in \%), CSIG, CBAK, and COVL in causal configuration, with IRM as the training objective and training length of 1s. The results of full-length processing (length generalization) and chunk processing (The utterance is split into 1s-long chunks) are reported. The non-overlapped and overlapped chunk processing are denoted by `-Seg' and '-Seg-O', respectively.  
    }
    \label{tab:magirm1s_causeg}
    \begin{tabular}{@{}cl||ccccc@{}}
        \toprule[1.25pt]
         &  & \multicolumn{5}{c}{\textbf{Metrics}}\\  
        Test Len. & Model & PESQ & ESTOI & CSIG & CBAK & COVL\\  
        
        \hline
        \hline
            
        \multirow{22}{*}{20s}
        & Noisy        & 1.91 & 51.42 & 2.36 & 1.82 & 1.72    \\
        \cdashline{2-7}
        & No-Pos         & 2.15 & 56.56 & 2.80 & 2.15 & 2.05  \\
        & No-Pos-Seg     & 2.34 & 61.90 & 3.01 & 2.35 & 2.26  \\
        & \textcolor{black}{No-Pos-Seg-O}  & \textcolor{black}{2.35}& \textcolor{black}{62.34} & \textcolor{black}{3.02} & \textcolor{black}{2.36}& \textcolor{black}{2.27} \\
        & RoPE         & 2.10 & 54.46 & 2.73 & 2.11 & 2.01  \\
        & RoPE-Seg     & 2.36 & 62.44  & 3.05 & 2.38 & 2.29  \\
        & \textcolor{black}{RoPE-Seg-O}   & \textcolor{black}{2.37} & \textcolor{black}{62.81}  & \textcolor{black}{3.06} & \textcolor{black}{2.38} & \textcolor{black}{2.30}  \\
        & T5-Bias      & 2.35 & 61.67 & 3.06 & 2.36 & 2.28  \\
        & T5-Bias-Seg      & 2.38 & 62.63 & 3.06 & 2.39 & 2.30  \\
        & \textcolor{black}{T5-Bias-Seg-O}   & \textcolor{black}{2.39} & \textcolor{black}{62.95} & \textcolor{black}{3.07} & \textcolor{black}{2.40} & \textcolor{black}{2.31}  \\
        & TISA         & 2.41 & 63.52 & 3.11 & 2.41 & 2.34  \\
        & TISA-Seg     & 2.36 & 62.46 & 3.03 & 2.37 & 2.28  \\
        & \textcolor{black}{TISA-Seg-O}   & \textcolor{black}{2.37} & \textcolor{black}{62.84} & \textcolor{black}{3.04} & \textcolor{black}{2.38} & \textcolor{black}{2.29}  \\
        & DA-Bias   & 2.44 & 63.32 & 3.16 & 2.43 & 2.38  \\
        & DA-Bias-Seg   & 2.38 & 62.56 & 3.05 & 2.37 & 2.29  \\
        & \textcolor{black}{DA-Bias-Seg-O}  & \textcolor{black}{2.39} & \textcolor{black}{62.89} & \textcolor{black}{3.06} & \textcolor{black}{2.37} & \textcolor{black}{2.30}  \\
        & \textcolor{black}{KERPLE}       & 2.45 & 64.13 & 3.17 & 2.45 & 2.40  \\
        & \textcolor{black}{KERPLE-Seg}   & 2.36 & 62.37 & 3.05 & 2.37 & 2.29  \\
        & \textcolor{black}{KERPLE-Seg-O}   & \textcolor{black}{2.37} & \textcolor{black}{62.78} & \textcolor{black}{3.06} & \textcolor{black}{2.37} & \textcolor{black}{2.30}  \\
        & LearnLin     & \textbf{2.48} & \textbf{64.59} & \textbf{3.19} & \textbf{2.48} & \textbf{2.43}  \\
        & LearnLin-Seg & 2.37 & 62.58 & 3.05 & 2.38 & 2.30  \\
        & \textcolor{black}{LearnLin-Seg-O} & \textcolor{black}{2.38} & \textcolor{black}{62.97} & \textcolor{black}{3.06} & \textcolor{black}{2.38} & \textcolor{black}{2.31}  \\
        \hline
        \toprule[1.25pt]
    \end{tabular}
    \vspace{-2.0em}
\end{table}